\documentclass[12pt,letterpaper]{article}

\usepackage{amsmath,amssymb,calc}
\usepackage{graphicx}
\usepackage{color}

\newcommand\be{\begin{equation}}
\newcommand\ee{\end{equation}}
\newcommand{\bea}{\begin{eqnarray}}
\newcommand{\eea}{\end{eqnarray}}

\newcommand{\nn}{\nonumber}
\newcommand{\pd}{\partial}

\def\id{\protect{{1 \kern-.28em {\rm l}}}}

\def\1{^{(1)}}
\def\0{^{(0)}}
\def\2{^{(2)}}

\def\id{\protect{{1 \kern-.28em {\rm l}}}}

\let\non\nonumber

\begin{document}

\begin{titlepage}
\begin{center}
\hfill \\

\vskip 2cm
{\Large \bf Glueball Spectrum in a Gauge Theory \vspace{0.2cm}\\ with Two Dynamical Scales}

\vskip 1.5 cm
{\bf  Lilia Anguelova${}^a$\footnote{languelova@perimeterinstitute.ca}, Peter Suranyi${}^b$\footnote{peter.suranyi@gmail.com} and L.C.R. Wijewardhana${}^b$\footnote{rohana.wijewardhana@gmail.com}\\
\vskip 0.5cm  {\it ${}^a$ Perimeter Institute for Theoretical Physics, Waterloo, ON N2L 2Y5, Canada\\ ${}^b$ Dept. $\!$of Physics, University of Cincinnati,
Cincinnati, OH 45221, USA}\non\\}

\end{center}
\vskip 2 cm
\begin{abstract}
We investigate the glueball spectrum of a strongly coupled gauge theory with two dynamical scales. The main tool is the use of the gauge/gravity duality. The model we study has a known graviational dual, which arises from a type IIB D-brane configuration. It exhibits two dynamical scales, separated by a nearly conformal region. Thus, it is of great interest for the study of walking in gauge theories. By using the gravitational description, we are able to compute analytically the glueball mass spectrum in a certain range of the parameter space of the theory. Within that range, we do not find a light state that could be associated with a slight breaking of conformal invariance. Finally, we show that in this model there can be an order of magnitude hierarchy between the scale of confinement, given by the lowest glueball mass, and the scale of chiral symmetry breaking, determined by the lowest vector-meson mass.
\end{abstract}
\end{titlepage}

\tableofcontents

\section{Introduction}

Gauge theories play an indispensable role in our understanding of nature. Yet, it is rather challenging to obtain information about their non-perturbative regime with standard field theory techniques. In recent years, a completely new tool for the study of this regime has emerged. This is the gauge/gravity duality, which maps strongly-coupled field theoretic problems into (almost) classical computations in a gravitational theory in a different number of dimensions. We will use this new method to investigate a particular class of gauge theories with two dynamical scales, which have attracted a great deal of attention lately. More precisely, these are theories with a gauge coupling that runs slowly within the energy range between the two scales. Such walking gauge theories have been thought to have relevance for models of dynamical electroweak symmetry breaking \cite{WalkT}. Regardless of that, however, they are an interesting and still poorly understood class of strongly-coupled gauge theories that is the simplest conceptual generalization of single-scale theories like QCD. Thus, their investigation can be viewed as a first step towards the study of the much more general case of gauge theories with multiple dynamical scales.

In the past, walking gauge theories have been analyzed via crude analytical approximations, like solving the Dyson-Schwinger gap equations with certain truncations of the integration kernel \cite{WalkT, CG}. More recent investigations in this area involved the extensive use of lattice gauge theory techniques to simulate slow running, induced by the screening of the gauge force that results from the inclusion of a large number of fermion flavors \cite{Lattice}. The initial analytic studies indicated that slow running of the gauge coupling would enhance the anomalous dimensions of the fermion bilinear operators. Furthermore, it was realized that slow running and enhanced anomalous dimensions are natural consequences of the gauge theory having an approximate fixed point, dominating the infrared dynamics, while being asymptotically free in the UV \cite{IRfpt}.

Such theories, with approximate infrared fixed-point behavior, exhibit approximate conformal symmetry due to the slow variation of the running gauge coupling over a large energy range. Dynamical chiral symmetry breaking would break this conformal symmetry, including scale invariance, spontaneously. The spontaneous breaking of exact scale invariance gives rise to a zero mass Nambu-Goldstone boson termed the dilaton, which couples to the trace of the energy momentum tensor of the theory. Naturally then, it was conjectured that the spontaneous breaking of approximate scale invariance in a walking theory would give rise to a pseudo Nambu-Goldstone boson, a dilaton whose mass is parametrically lighter than the typical mass scales of the theory, as measured by the vector boson masses \cite{ConjD}. There is a great deal of confusion in the literature about the validity of this conjecture with some authors disagreeing with its existence \cite{AgainstD}, while others agreeing with varying degrees of enthusiasm \cite{ForD}.

A definitive statement about the existence of such a light dilaton state cannot be made without employing a more rigorous non-perturbative study of a walking theory. Lattice gauge computer simulations would be one way to achieve this goal. Another approach is to study such a model using the gauge/gravity duality method. The work \cite{NPP} recently constructed a gravity dual of a walking theory by considering a background sourced by a set of D5 branes, which is a deformation of the famous Maldazena-Nunez solution \cite{MN}. The four-dimensional walking gauge theory arises from the world-volume effective action of the D5 branes, wrapped on a certain 2-cycle. This gravitational background encodes the color degrees of freedom. Flavor degrees of freedom can be added by embedding additional probe branes in the background. In order to obtain a gravity dual, that captures chiral symmetry breaking, \cite{LA} introduced a U-shaped embedding of D7-$\overline{{\rm D7}}$ branes in the vein of the Sakai-Sugimoto model \cite{SS}. In two subsequent papers \cite{ASW,ASW2}, we analyzed the vector and scalar meson spectra of the flavor sector in \cite{LA}. The results did not contain a parametrically light scalar.\footnote{We should note that the later work \cite{CLV} finds an instability, due to a tachyonic mode, in the scalar meson spectrum of a rather similar model. We would like to underline, though, that there is a crucial difference between that model and ours. Namely, we have a UV cutoff determined by the upper end of the walking region. This is due to our physical perspective, explained in the next paragraph. With this cutoff, we do not find any instability, as shown in \cite{ASW2}.} However, as pointed out in \cite{ASW2}, if a dilaton exists in this model, it would naturally reside in the color sector. So here we will investigate the scalar spectrum of the color background of \cite{NPP} that underlies our model.

We should point out that  in this paper we have the same rational as in \cite{ASW,ASW2}. Namely, we view this model as an effective description, valid only below a certain physical scale, much like the four-Fermi model. In phenomenological applications, the appropriate UV completion should be able to encode extended technicolor. This is an interesting open question, which we do not address in the present paper.\footnote{For some progress in that direction see \cite{CGNPR}.} The above perspective leads to important differences from the work of \cite{ENP}, which viewed a background that is a modification of \cite{NPP} as a full UV complete description.\footnote{Later on, we will see a more technical difference as well. We thank M. Piai for a useful discussion on these issues.} The numerical study \cite{ENP} found evidence for the existence of a parametrically light state in the glueball spectrum.
However, the numerical method is not transparent enough to allow for the identification of the light state with the dilaton. So here we will investigate this issue for our model with analytical means, in order to achieve better understanding.

In addition to the purely theoretical interest in this gauge theory problem, such studies may also have phenomenological relevance. The recent discovery of the 125 GeV Higgs-like state at the LHC has severely constrained the dynamical models of electro-weak symmetry breaking. In QCD-like dynamical models, the composite scalar excitations are generically as massive as the vector meson ones. However, in principle, the situation could be rather different in a walking gauge theory, where the dynamics is much more involved than in a QCD-like theory. Therefore, it is important to explore whether a walking theory can have a composite light scalar mode, whose behavior could mimic the Standard Model Higgs. The natural candidate is the dilaton discussed above. The reason is that since it would arise as a pseudo Nambu-Goldstone boson, its mass would be naturally low compared to the rest of the spectrum. Of course, this by itself would not be enough. Such a state would also have to have appropriate couplings to other fields, to be consistent with current data.

Recent works have studied this possibility. See for example \cite{Dil}. Also, the couplings of such a dilaton were worked out by \cite{GGS}, assuming that the Standard Model is embedded in a conformal field theory. Other options have been investigated in a number of papers. For example, the possibility that the 125 GeV state is a composite psedo-scalar was discussed by \cite{Imp}. A light composite Higgs was also considered in \cite{LCH}. And in \cite{PhenImp} a variety of phenomenological studies was performed to explore different composite particles, generically called Higgs impostors. Clearly, more experimental data is needed in order to determine whether the observed state corresponds to a fundamental scalar or any of the above composite possibilities.

Regardless of any phenomenological motivation though, as we have already pointed out, the presence or absence of a dilaton in a walking gauge theory is an interesting open problem regarding the strongly-coupled regime. In this paper, we will address it for a particular model, whose gravity dual is based on the solution of \cite{NPP}, as explained above. To that end, we will compute analytically the spectrum of scalar glueball fluctuations around that background in a certain subspace of the parameter space of the theory. It turns out that, in that range, the spectrum does not contain a parametrically light state (dilaton).
We also show that, for a certain parameter range, there can be an order of magnitude hierarchy between the vector meson and glueball spectra. This hierarchy measures the ratio between the scales of confinement and chiral symmetry breaking.

Since in \cite{ASW} we found that the mass of the rho meson (the lightest vector meson) in our model is of order $\gtrapprox 5$ TeV for phenomenologically allowed values of the S-parameter, the above hierarchy implies that the lightest glueball has a mass in the range $0.5$ - $2$ TeV. It should be noted that, in principle, the mass of such a particle could be reduced, when taking into account its couplings to Standard Model fields \cite{FFS}. It would be interesting to explore whether that could happen in our model, as in such a case this composite scalar would have mass in just the right range compared to the Higgs-like state observed at the LHC. We underline again, though, that this scalar is not a dilaton.

\section{Gravitational dual}
\setcounter{equation}{0}

The gravitational dual of interest for us is an ${\cal N} = 1$ solution of the type IIB equations of motion. It arises from a stack of $N_c$ D5 branes wrapped on a two-sphere and has nontrivial RR 3-form flux, string dilaton and ten-dimensional metric. The latter is characterized by two additional parameters $c$ and $\alpha$. In fact, the solution is obtained as an expansion in $1/c$, where $c>\!\!>1$. To leading order, the 10d metric is \cite{NPP}:\footnote{Here we have corrected a typo/mistake in eq. (36) of \cite{NPP}, which has no effect on the considerations of \cite{LA,ASW,ASW2}.}
\bea \label{BM}
ds^2 \!&=& \!A \left[ dx_{1,3}^2 + \frac{cP_1' (\rho)}{8} \left( 4 d\rho^2 + (\omega_3 + \tilde{\omega}_3)^2 \right) \right. \nn \\
&+& \!\left. \frac{c\,P_1 (\rho)}{4} \left( \frac{1}{\coth(2\rho)} d\Omega_2^2 + \coth(2\rho) d\tilde{\Omega}_2^2 + \frac{2}{\sinh ( 2\rho)} (\omega_1 \tilde{\omega}_1 - \omega_2 \tilde{\omega}_2) \right) \right]\!.
\eea
where
\be \label{AP1}
A=\left( \frac{3}{c^3 \sin^3 \alpha} \right)^{1/4} , \quad  P_1' (\rho) =\frac{\pd P_1 (\rho)}{\pd \rho} \,\, , \quad P_1 (\rho) = \left( \cos^3 \alpha + \sin^3 \alpha \left( \sinh (4 \rho) - 4 \rho \right) \right)^{1/3} \, ,
\ee
\bea \label{omegatilde}
\tilde{\omega}_1 &=& \cos \psi d\tilde{\theta} + \sin \psi \sin \tilde{\theta} d \tilde{\varphi} \,\, , \hspace*{2cm} \omega_1 = d \theta \,\, , \nn \\
\tilde{\omega}_2 &=& - \sin \psi d\tilde{\theta} + \cos \psi \sin \tilde{\theta} d \tilde{\varphi} \,\, , \hspace*{1.6cm} \omega_2 = \sin \theta  d \varphi \,\, , \nn \\
\tilde{\omega}_3 &=& d \psi + \cos \tilde{\theta} d \tilde{\varphi} \,\, , \hspace*{3.7cm} \omega_3 = \cos \theta d \varphi
\eea
and
\be
d\tilde{\Omega}_2^2 = \tilde{\omega}_1^2 + \tilde{\omega}_2^2 \,\, , \hspace*{2cm} d\Omega_2^2 = \omega_1^2 + \omega_2^2 = d \theta^2 + \sin^2 \theta d \varphi^2 \,\, .
\ee

In the walking region, $\rho$ is always of order 1 or larger. Hence, $\coth(2\rho) \approx 1$ while $\frac{1}{\cosh(2\rho)}$ is negligible. In addition, the walking region is characterized by:
\be
\beta \equiv \sin^3 \alpha <\!\!< 1 \, .
\ee
As a result, to leading order in small $\beta$, one can use the approximations:
\be \label{approxP}
P_1 = 1 \qquad , \qquad P_1' = \frac{2}{3} \beta e^{4 \rho} \, .
\ee
Therefore, the metric (\ref{BM}) simplifies to:
\be \label{leadMetric}
ds^2_{{\rm walk}} = A \left[ \eta_{\mu \nu} dx^{\mu} dx^{\nu} + \frac{c}{12} \beta e^{4 \rho} \left( 4 d\rho^2 + \left( \omega_3 + \tilde{\omega}_3 \right)^2 \right) + \frac{c}{4} \left( d \Omega_2^2 + d \tilde{\Omega}_2^2 \right) \right] .
\ee

We will also need the RR 3-form field strength $F^{RR}_3$ (see eq. (6) of \cite{NPP}):
\bea \label{F3}
F_3^{RR} &=& \frac{N_c}{4} \bigg[ - (\tilde{\omega}_1 + b d \theta) \wedge (\tilde{\omega}_2 - b \sin \theta d \varphi) \wedge (\tilde{\omega}_3 + \cos \theta d\varphi) + \nn \\
&+& b' d\rho \wedge (- d\theta \wedge \tilde{\omega}_1 + \sin \theta d \varphi \wedge \tilde{\omega}_2) + (1-b^2) (\sin \theta d \theta \wedge d\varphi \wedge \tilde{\omega}_3) \bigg] \,\,\, ,
\eea
where
\be
b (\rho) = \frac{2 \rho}{\sinh(2 \rho)} \,\,\, , \nn
\ee
as well as the string dilaton $\phi$:
\be \label{strdil}
e^{\phi} = \left( \frac{3}{c^3 \sin^3 \alpha} \right)^{1/4} \, .
\ee

The metric (\ref{BM}), RR 3-form (\ref{F3}) and dilaton (\ref{strdil}) together solve the type IIB supergravity equations of motion (to leading order \cite{NPP}). To be more explicit, the relevant field equations are:
\be \label{EoM1}
R_{AB} = \frac{1}{2} \pd_A \phi \pd_B \phi + \frac{1}{4} \left( F_{ACD} F^{CD}{}_B  - \frac{1}{12} g_{AB} F^2_3 \right)
\ee
\be \label{EoM2}
\pd_A \left( \sqrt{-\det g} \, e^{\phi} \, F_3^{ABC} \right) = 0
\ee
\be \label{EoM3}
\nabla^2 \phi = \frac{e^{\phi}}{12} \, F_3^2 \,\, ,
\ee
where the indices $A,B,C,D$ run over all 10 dimensions and $F_3 \equiv F_3^{RR}$. Also, the 3-form satisfies the usual Bianchi identity:
\be
\pd_{[A} F_{BCD]} = 0 \,\, .
\ee

In the following, we will be studying fluctuations of the above solution, in order to obtain the corresponding glueball spectrum. Since a main goal for us is to see whether this spectrum contains a possible dilaton, we will also address the coupling of those fluctuations to a supersymmetric D5 probe embedded in (\ref{BM}). The latter is needed in order to determine the coupling of a prospective dilaton with the color gauge field strength, corresponding to the supersymmetric dual background. Finally, to preserve susy, the D5 probe has to wrap the only supersymmetric cycle in the above gravity background, namely the two-dimensional surface defined by \cite{NPP}:
\be \label{sigma2}
\Sigma_2 \quad : \quad \theta = \tilde{\theta} \, , \,\, \varphi = 2 \pi - \tilde{\varphi} \, , \,\, \psi = \pi \, .
\ee
So the six-dimensional worldvolume of the D5 probe has to be along the 4d Minkowski space, parametrized by $x^{\mu}$, and along the surface $\Sigma_2$.

\section{Fluctuations}
\setcounter{equation}{0}

The fluctuations of the background are of the form:
\bea \label{fluct}
\phi &=& \phi_w + \delta \phi \, , \nn \\
g_{AB} &=& g_{AB}^w + \delta g_{AB} \, , \nn \\
F_3 &=& F_3^w + \delta F_3 \, ,
\eea
where $w$ denotes the walking background solution [i.e., (\ref{BM}), (\ref{F3}), (\ref{strdil})] and the fluctuations $\delta \phi$, $\delta g_{AB}$, $\delta F_3$ have to satisfy the linearized equations of motion [i.e., the linearization of (\ref{EoM1}) - (\ref{EoM3})]. As before, we would like to study fluctuations of the form:
\be \label{fluct2}
\delta \phi = \delta \phi (x^{\mu}, \rho) \quad , \quad \delta g_{AB} = \delta g_{AB} (x^{\mu}, \rho) \quad , \quad \delta F_3 = \delta F_3 (x^{\mu}, \rho) \, .
\ee
Note also that we are only interested in scalar fluctuations. By that we mean fields, which are scalars with respect to the symmetry of the 4d Minkowski space with coordinates $x^{\mu}$. So, generically, the number of scalar fluctuations is significant. To perform the proper counting, let us first write schematically the general form of the 10d background metric we have:
\be
ds^2_w = A \eta_{\mu \nu} dx^{\mu} dx^{\nu} + f(\rho) d\rho^2 + g_{ab}^w dy^a dy^b \, ,
\ee
where $A=const$, $f(\rho)$ is some function of $\rho$ and the coordinates $y^a$ are the set of internal angles, i.e. $\{ y^a \} = \{ \varphi, \theta, \tilde{\varphi}, \tilde{\theta}, \psi \}$. The most general perturbation of the above metric is of the form $ds^2 = ds^2_w + ds^2_{pert}$ with:
\bea
ds^2_{pert} &=& (\delta g_{\mu \nu}) dx^{\mu} dx^{\nu} + 2 (\delta g_{\mu \rho}) dx^{\mu} d\rho + (\delta g_{\rho \rho}) d\rho^2 + (\delta g_{ab}) dy^a dy^b + 2 (\delta g_{\mu a}) dx^{\mu} dy^a \nn \\
&+& 2 (\delta g_{a \rho}) dy^a d\rho \,\, .
\eea
One can easily count that the number of components of the symmetric 10d matrix $\delta g_{AB}$, with $\{ A,B \} = \{ x^{\mu}, \rho, y^a \}$, is equal to $55$, as it should be. Now, the perturbations $\delta g_{\mu a}$ give five four-dimensional vectors (for each of the five values of $a$ we have a vector in the 4d space with coordinates $x^{\mu}$). Similarly, $\delta g_{\mu \rho}$ is also a vector perturbation in four dimensions. However, each of the components of $\delta g_{\rho \rho}$, $\delta g_{ab}$ and $\delta g_{a \rho}$ is a scalar in 4d. This gives in total $1+ 15 + 5 = 21$ scalars. Furthermore, the 4d fluctuation $\delta g_{\mu \nu}$ can be decomposed into trace and traceless parts; the traceless part represents the 4d tensor modes, while the trace is another scalar. So there are $22$ scalars in total, coming from the metric. In addition, there are also the fluctuations of $\phi$ and $F_3$. Clearly, this is quite a lot to handle, if all of those scalars are coupled to each other via the equations of motion. Fortunately, it turns out that there is a consistent truncation of type IIB supergravity to an effective 5d theory, which is much simpler to analyze then the full fluctuated 10d system. The space-time of the five-dimensional theory is the $(x^{\mu},\rho)$ space and the field content is a 5d metric and six scalars. Recall that a consistent truncation means that every solution of the field equations of the 5d theory is guaranteed to also give a solution of the full ten-dimensional equations of motion. Furthermore, comparing with \cite{ENP}, the consistent truncation we will consider should be enough for our purposes (i.e., it should contain the relevant light mode).

\subsection{Consistent truncation: generalities}

The field content of type IIB supergravity is the following: 10d metric $g_{AB} (x^M)$, string dilaton $\phi(x^M)$, NS 3-form field $H_3 (x^M)$, RR scalar $C(x^M)$, RR 3-form $F_3 (x^M)$ and RR 5-form $F_5 (x^M)$, where $A,B,M = 0,...,9$ are 10d indices. For the deformed Maldacena-Nunez solutions, that encompass the background of interest for us, three of those fields are identically zero: $C(x^M) = 0$, $H_3 (x^M) = 0$ and $F_5 (x^M) = 0$. The bosonic action for the remaining fields is:
\be \label{IIBAction}
S_{IIB} = \frac{1}{G_{10}} \int d^{10}x \sqrt{-\det g} \left[ R - \frac{1}{2} (\pd \phi)^2 - \frac{e^{\phi}}{12} F_3^2 \right] \, .
\ee

Let us consider a 10d metric of the form:
\bea \label{MNMetric}
ds^2 &=& g_{IJ} dx^I dx^J +  e^{2f_1} (d\theta^2 + \sin^2 \theta d\varphi^2) + \frac{e^{2 f_2} }{4} \left[ \left( \tilde{\omega}_1 + a d\theta \right)^2 + \left( \tilde{\omega}_2 - a \sin \theta d \varphi \right)^2 \right] \nn \\
&+& \frac{e^{2f_3}}{4} \left( \tilde{\omega}_3 + \cos \theta d\varphi \right)^2 ,
\eea
where $x^I = \{ x^{\mu}, \rho\}$ and the four quantities $f_1$, $f_2$, $f_3$, $a$ are all functions of $x^I$ only, but not of the five angles $\varphi$, $\theta$, $\tilde{\varphi}$, $\tilde{\theta}$, $\psi$; here $\tilde{\omega}_{1,2,3}$ are as in (\ref{omegatilde}). Furthermore, the string dilaton is also a function
\be
\phi = \phi (x^{\mu}, \rho)
\ee
and the RR 3-form is of the form:
\bea \label{RRflux}
F_3^{RR} &=& \frac{N_c}{4} \bigg[ - (\tilde{\omega}_1 + b d \theta) \wedge (\tilde{\omega}_2 - b \sin \theta d \varphi) \wedge (\tilde{\omega}_3 + \cos \theta d\varphi) + \nn \\
&+& b' d\rho \wedge (- d\theta \wedge \tilde{\omega}_1 + \sin \theta d \varphi \wedge \tilde{\omega}_2) + (1-b^2) (\sin \theta d \theta \wedge d\varphi \wedge \tilde{\omega}_3) \bigg] \,\,\, ,
\eea
where $b = b(x^{\mu},\rho)$. Then, according to \cite{BHM}, the theory (\ref{IIBAction}) has a consistent truncation to a 5d theory for the metric $g_{IJ} (x^I)$ and the six scalars $f_1 (x^I)$, $f_2 (x^I)$, $f_3 (x^I)$, $\phi (x^I)$, $a(x^I)$, $b(x^I)$ with action obtained by substituting (\ref{MNMetric})-(\ref{RRflux}) into (\ref{IIBAction}) and integrating over the five internal angles. In fact, to obtain diagonal kinetic terms for the six 5d scalars, it is useful to rewrite (\ref{MNMetric}) in the form (in the notation of \cite{BHM}):
\bea \label{MNMetric2}
ds^2 &=& e^{2p -x} g_{IJ} dx^I dx^J + e^{x+g} (d\theta^2 + \sin^2 \theta d\varphi^2) + e^{x-g} \left[ \left( \tilde{\omega}_1 + a d\theta \right)^2 + \left( \tilde{\omega}_2 - a \sin \theta d \varphi \right)^2 \right] \nn \\
&+& e^{-6p-x} \left( \tilde{\omega}_3 + \cos \theta d\varphi \right)^2 ,
\eea
where $p=p(x^I)$, $x=x(x^I)$, $g=g(x^I)$. Clearly, (\ref{MNMetric}) and (\ref{MNMetric2}) are related by the change of variables $\{f_1, f_2, f_3\} \rightarrow \{ p, x, g \}$ together with a rescaling of the 5d metric, which is perfectly allowed since $g_{IJ} (x^I)$ is arbitrary. Now, the set of fields $\Phi^i = \{g,p,x,\phi,a,b\}$ and $g_{IJ}$ is described by the 5d action \cite{BHM}:
\be \label{5dAc}
S = \int d^5 x \sqrt{-\det g} \left[ \frac{R}{4} - \frac{1}{2} G_{ij} \pd \Phi^i \pd \Phi^j - V(\{ \Phi^i \}) \right] \, ,
\ee
where the non-linear sigma model metric $G_{ij}$ is diagonal and has the following components:
\be \label{SigmaMM}
G_{pp} = 6 \,\,\, , \,\,\, G_{xx} = 1 \,\,\, , \,\,\, G_{gg} = \frac{1}{2} \,\,\, , \,\,\, G_{\phi \phi} = \frac{1}{4} \,\,\, , \,\,\, G_{aa} = \frac{e^{-2g}}{2} \,\,\, , \,\,\, G_{bb} = \frac{N_c^2 e^{\phi - 2x}}{32} \,\,\, ,
\ee
while the potential $V(\{ \Phi^i \})$ is \cite{BHM,DE}:\footnote{Here we have arranged the terms in $V(\Phi^i)$ as in \cite{DE} for easier comparison with \cite{ENP}.}
\bea \label{pot}
V(\{ \Phi^i \}) \!\!\!&=& \!\!\!\frac{e^{-2g -4 (p+x)}}{128} \bigg[ 16 \left\{ a^4 + 2 \left( (e^g - e^{6p+2x})^2 - 1 \right) a^2 + e^{4g} - 4 e^{g+6p+2x} (1+e^{2g}) + 1 \right\} \nn \\
&+& \!\!\!e^{12p+2x+\phi} \left\{ 2e^{2g} (a-b)^2 + e^{4g} + (a^2 - 2ab + 1)^2 \right\} N_c^2 \bigg] \,.
\eea

In \cite{BHM} it was shown that every solution of the field equations of (\ref{5dAc}) satisfies automatically the full 10d equations of motion of (\ref{IIBAction}). Therefore, one can study consistently a subset of all fluctuations (\ref{fluct}) by expanding around a ($\rho$-dependent walking) background solution as:
\be
g_{IJ} (\rho, x^{\mu}) = g_{IJ}^{w} (\rho) + \delta g_{IJ} (\rho, x^{\mu}) \quad , \quad \Phi^i (\rho, x^{\mu}) = \Phi^i_w (\rho) + \delta \Phi^i (\rho, x^{\mu})
\ee
and investigating the equations of motion for $\delta g_{IJ}$ and $\delta \Phi^i$ that follow from (\ref{5dAc}).

All of this is useful for us because the walking background of \cite{NPP} has a 10d metric of the form:
\bea \label{MetricCTr}
ds^2 \!&=& \!e^{2\hat{f}} \bigg\{ dx_{1,3}^2 + e^{2k} d\rho^2 + e^{2h} (d\theta^2 + \sin^2 \theta d\varphi^2) + \frac{e^{2 \hat{g}} }{4} \left[ \left( \tilde{\omega}_1 + a d\theta \right)^2 + \left( \tilde{\omega}_2 - a \sin \theta d \varphi \right)^2 \right] \nn \\
&+& \!\frac{e^{2k}}{4} \left( \tilde{\omega}_3 + \cos \theta d\varphi \right)^2 \bigg\} \,\, ,
\eea
which is a special case of (\ref{MNMetric2}), as well as 3-form $F_3$ precisely of the form (\ref{RRflux}). Hence we can use the 5d action (\ref{5dAc}) to study the fluctuations we are interested in.

\subsection{Application to walking background}

In the walking region, we have (to leading order) \cite{NPP}:
\bea \label{FuncWp}
&& e^{\phi_w} = A = const \quad , \quad 2 \hat{f}_w = \phi_w \quad , \quad e^{2k_w} = \frac{c P'_1}{2} \quad , \nn \\
&& e^{2h_w} = \frac{c P_1}{4 \coth (2\rho)} \quad , \quad e^{2\hat{g}_w} = c P_1 \coth (2\rho) \quad , \nn \\
&& a_w = \frac{1}{\cosh (2 \rho)} \approx 0 \quad , \quad b_w = \frac{2\rho}{\sinh (2\rho)} \approx 0 \quad ,
\eea
where $A$ and $P_1$ are as in (\ref{AP1}). In particular, the background RR 3-form is (to leading order):
\be
F_3^{RR} \approx \frac{N_c}{4} \left( \omega_1 \wedge \omega_2 - \tilde{\omega}_1 \wedge \tilde{\omega}_2 \right) \wedge \left( \omega_3 + \tilde{\omega}_3 \right) \, ;
\ee
see also equation (40) of \cite{NPP}. Using (\ref{approxP}) and $\coth(2\rho) \approx 1$, we then obtain the following background functions:
\bea \label{BkgrF}
&& e^{\phi_w} = e^{2 \hat{f}_w} = A \quad , \quad e^{2k_w} = \frac{c}{3} \beta e^{4\rho} \quad , \quad e^{2h_w} = \frac{c}{4} \quad , \nn \\
&& e^{2 \hat{g}_w} = c \quad , \quad a_w = 0 \quad , \quad b_w = 0 \quad .
\eea
Substituting these in (\ref{MetricCTr}) gives exactly our walking metric (\ref{leadMetric}), which is consistent (as it should be) with the walking expression in eq. (39) of \cite{NPP} (modulo the typo they have regarding the overall constant multiplying the metric).

Comparing (\ref{MNMetric2}) and (\ref{MetricCTr}), we find:
\be
e^{x+g} = e^{2h+2\hat{f}} \quad , \quad e^{x-g} = \frac{e^{2\hat{g} + 2 \hat{f}}}{4} \quad , \quad e^{-6p -x} = \frac{e^{2k + 2 \hat{f}}}{4} \quad .
\ee
Hence (\ref{BkgrF}) implies that in the walking background:
\be \label{Sol1}
e^{- 6 p_w} = \frac{A^3 c^3}{4^3} \frac{\beta}{3} e^{4\rho} \quad , \quad e^{2 x_w} = \frac{A^2 c^2}{16} \quad , \quad e^{2 g_w} = 1 \quad .
\ee
Plus, of course, we still have:
\be \label{Sol2}
e^{\phi_w} = A \quad , \quad a_w =0 \quad , \quad b_w = 0 \quad .
\ee
Therefore, (\ref{Sol1})-(\ref{Sol2}) is the solution for the six scalars $\Phi^i$, around which we will expand in small fluctuations, namely:
\bea \label{SmallFl}
&& p= p_w + \delta p(\rho, x^{\mu}) \quad , \quad x = x_w + \delta x(\rho, x^{\mu}) \quad , \quad g = \delta g(\rho , x^{\mu}) \nn \\
&& \phi = \phi_w + \delta \phi (\rho, x^{\mu}) \quad , \quad a = \delta a(\rho, x^{\mu}) \quad , \quad b = \delta b (\rho, x^{\mu}) \quad .
\eea
Note also that, in the walking background (\ref{MetricCTr}), the 5d part of the metric (\ref{MNMetric2}), i.e. $e^{2p-x} g_{IJ} dx^I dx^J$, acquires the form $e^{2\hat{f}} dx_{1,3}^2 + e^{2(\hat{f} + k)} d\rho^2$. In other words, the background 5d metric $g_{IJ}$, around which we need to expand, is of the form:\footnote{Recall that $R$ in (\ref{5dAc}) is the scalar curvature of the 5d metric $g_{IJ}$.}
\bea \label{5dBM}
g_{IJ}^w dx^I dx^J &=& e^{-2p_w+x_w} \left( e^{2\hat{f}_w} \eta_{\mu \nu} dx^{\mu} dx^{\nu} + e^{2(\hat{f}_w+k_w)} d\rho^2 \right) \nn \\
&=& \frac{A^3 c^2}{16} \left( \frac{\beta}{3} \right)^{1/3} \!e^{\frac{4\rho}{3}} \left( \eta_{\mu \nu} dx^{\mu} dx^{\nu} + \frac{c}{3} \beta e^{4\rho} d\rho^2 \right) \, .
\eea
In addition to (\ref{SmallFl}), we also need to expand:
\be \label{5dMetricFl}
g_{IJ} = g_{IJ}^w + \delta g_{IJ} (\rho, x^{\mu}) \, .
\ee
The field equations for all fluctuations in (\ref{SmallFl}) and (\ref{5dMetricFl}) follow from the action (\ref{5dAc}).

Note that, to second order in perturbations, the kinetic terms in (\ref{5dAc}) acquire the form:
\bea \label{KinT}
{\cal L}_{kin} =  -\frac{1}{2} G_{ij} \pd \Phi^i \pd \Phi^j &=& - 3 \pd_I (\delta p) \pd^I (\delta p) - 6 \pd_{\rho} (p_w) \pd^{\rho} (\delta p) - 3 \pd_{\rho} (p_w) \pd^{\rho} (p_w) \nn \\
&-& \frac{1}{2} \pd_I (\delta x) \pd^I (\delta x) - \frac{1}{4} \pd_I (\delta g) \pd^I (\delta g) - \frac{1}{8} \pd_I (\delta \phi) \pd^I (\delta \phi) \nn \\
&-& \frac{1}{4} \pd_I (\delta a) \pd^I (\delta a) - \frac{N_c^2}{4 A c^2} \pd_I (\delta b) \pd^I (\delta b) \,\, ,
\eea
where again $I = \{ \mu, \rho \}$ and we have used (\ref{Sol1})-(\ref{SmallFl}).

The work \cite{BHM} studies the field equations for an action of the form (\ref{5dAc}) with any potential $V (\Phi^i)$ and any sigma-model metric $G_{ij}$ and derives the linearized equations of motion in terms of certain gauge invariant variables.\footnote{Clearly, since the 5d space-time metric $g_{IJ}$ is dynamical, the relevant gauge transformations are diffeomorphisms.} Those equations of motion are then used in the numerical study of \cite{ENP}. However, the direct numerical approach lacks transparency with respect to the physical meaning of the results. In order to gain better understanding, we will investigate these equations analytically for our model. This will allow us to see interesting new features. Also, we will find a different outcome regarding the existence of a dilaton in our case.

Before turning to that, let us note that in our case the gauge-invariant variables of \cite{BHM} simplify  significantly. The reason is that only one of the six scalars $p,x,g,\phi,a,b$ has a non-constant background value, namely $p$ whose background profile $p_w$ is a function of $\rho$; see (\ref{Sol1})-(\ref{Sol2}). This is important since a scalar transforms under infinitesimal diffeomorphisms (the gauge transformations of gravity) only if it has a nontrivial background profile. Indeed, recall that under the transformation:
\be \label{InfDiff}
x^I \quad \rightarrow \quad \tilde{x}^I = x^I + \xi^I \, ,
\ee
with $\xi^I$ infinitesimal, a scalar function $\Phi (x^I)$ transforms as:
\be
\Phi \quad \rightarrow \quad \tilde{\Phi} = \Phi + \xi^I \pd_I \Phi \, .
\ee
Now, decomposing $\Phi = \Phi_B + \delta \Phi$, where $\Phi_B$ is the background value and $\delta \Phi$ is a small fluctuation, we have to first order:
\be
\Phi_B + \delta \Phi \quad \rightarrow \quad \Phi_B + \delta \Phi + \xi^I \pd_I (\Phi_B) \, .
\ee
In other words, under the gauge transformation (\ref{InfDiff}), the fluctuation $\delta \Phi$ transforms (to first order) as:
\be \label{SGtr}
\delta_{GT} (\delta \Phi) = \xi^I \pd_I (\Phi_B) \, ,
\ee
where the subscript $GT$ in $\delta_{GT}$ stands for gauge transformation. So, whenever $\Phi_B = const$, the fluctuation $\delta \Phi$ is gauge invariant. Therefore, the five fluctuations $\delta x$, $\delta g$, $\delta \phi$, $\delta a$ and $\delta b$ are already gauge invariant.  So we only need to worry about the gauge-variance of the fluctuation $\delta p$. Interestingly, in terms of the original warp factors in (\ref{MetricCTr}), the only gauge-variant one is $k$; see (\ref{BkgrF}). However, this is the one warp factor, which does not affect the coupling to a susy D5 probe, since upon imposing (\ref{sigma2}) one has $\tilde{\omega}_3 + \cos \theta d\varphi = 0$. In other words, even if the solution we will find for the mode $\delta p$ turns out to contain a gauge artifact part, this will not affect the physical couplings we are interested in.

\subsection{Equations of motion}

The papers \cite{DE,BHM} derived the equations of motion for the six gauge-invariant scalar variables in the above set-up. To make use of their results, let us first write the background for the 5d metric $g_{I\!J}$, entering the action (\ref{5dAc}), in the form:
\be \label{BMAhat}
g_{IJ}^w dx^I dx^J = e^{2\hat{A}} \eta_{\mu \nu} dx^{\mu} dx^{\nu} + dz^2 \quad ,
\ee
where [see eq. (\ref{5dBM})]:
\be \label{Ahat}
e^{2\hat{A}} = e^{2 \hat{f}_w -2 p_w + x_w} = \frac{A^3 c^2}{16} \left( \frac{\beta}{3} \right)^{1/3} \!e^{\frac{4 \rho}{3}} \qquad {\rm and} \qquad dz^2 = \frac{A^3 c^2}{16} \left( \frac{\beta}{3} \right)^{1/3} \!e^{\frac{4 \rho}{3}} \,\frac{c}{3} \beta e^{4\rho} \,d\rho^2 \quad .
\ee
From now on, $\hat{A}$ is viewed as a function of $z$ via the second relation in (\ref{Ahat}). To be more explicit, we have
\be \label{zdef}
z = 3 \frac{\sqrt{A^3 c^3}}{32} \left( \frac{\beta}{3} \right)^{2/3} e^{\frac{8 \rho}{3}} \, .
\ee
Note that, ultimately, we want to keep all dependence on $\beta$ explicit. Hence, once we write down the equations of motion of \cite{DE,BHM}, we have to make another change of variables $z \rightarrow u$, where say $z = 3 \frac{\sqrt{A^3 c^3}}{32} \left( \frac{\beta}{3} \right)^{2/3} u^{2/3}$ in order to restore the factors of $\beta$. The latter transformation, in particular, means using
\be \label{Defu}
u=e^{4\,\rho}
\ee
as the radial variable.

Now, the linearized equations of motion that follow from (\ref{5dAc}), when expanding around a background metric of the form (\ref{BMAhat}), were derived in \cite{DE} following the work of \cite{BHM}. Those equations were obtained in terms of gauge-invariant fluctuations:
\be
\Phi^i \rightarrow \Phi_B^i + (\delta \Phi)_{gi}^i \, ,
\ee
where $\Phi_B^i$ are the background values of the original fields $\Phi^i = \{ g , p, x, \phi,a,b \}$. Since, in our case, the scalars $g$, $x$, $\phi$, $a$ and $b$ are already gauge invariant [see the discussion in and around (\ref{InfDiff})-(\ref{SGtr})], we have:
\be \label{FlVar}
(\delta g)_{gi} = \delta g \,\,\, , \,\,\, (\delta x)_{gi} = \delta x \,\,\, , \,\,\, (\delta \phi)_{gi} = \delta \phi \,\,\, , \,\,\, (\delta a)_{gi} = \delta a \,\,\, , \,\,\, (\delta b)_{gi} = \delta b \,\,\, .
\ee
The only scalar with a non-constant background is $p = p_w + \delta p (\rho, x^{\mu})$. Hence, the corresponding gauge-invariant fluctuation is \cite{BHM}:
\be
(\delta p)_{gi} = \delta p - \frac{(\pd_z p_w)}{6 (\pd_z \hat{A})} h \,\, ,
\ee
where $h$ is a certain metric fluctuation.\footnote{More precisely, up to a numerical constant, $h$ is the fluctuation $\zeta$ in (\ref{Metric1stOr}).}
According to \cite{DE}, the equations of motion are:
\bea \label{EOM}
&&\left[ D_z^2 + 4 \hat{A}' D_z + e^{-2\hat{A}} \pd_{\mu} \pd^{\mu} \right] \!(\delta \Phi)^i_{gi} - \nn \\
&&\left[ V^i{}_{|j} - {\cal R}^i{}_{kjl} \Phi'^k \Phi'^l + \frac{4 \left( \Phi'^i V_j + V^i \Phi'_j \right)}{3 \hat{A}'} + \frac{ 16 V \Phi'^i \Phi'_j }{9 \hat{A}'^2} \right] (\delta \Phi)^j_{gi} = 0 \,\,\, ,
\eea
where $' \equiv \pd_z$ and also:
\bea
&&D_z (\delta \Phi)^i = \pd_z (\delta \Phi)^i + \Gamma^i_{jk} \Phi_B'^j (\delta \Phi)^k \quad , \nn \\
&&V_i = \frac{\pd V}{\pd \Phi^i} \quad , \quad V^i = G^{ij} V_j \quad , \quad \Phi_i = G_{ij} \Phi^j \quad , \nn \\
&&V^i{}_{|j} = \pd_j V^i + \Gamma^i_{jk} V^k
\eea
with $\Gamma^i_{jk}$ and ${\cal R}^i{}_{kjl}$ being the Christoffel symbols and the Riemann curvature, respectively, of the sigma model metric $G_{ij}$ in (\ref{SigmaMM}).

In principle, the field equations (\ref{EOM}) look rather complicated. However, in our case they simplify significantly. To see that, let us first write down the Christoffel symbols for the metric (\ref{SigmaMM}):
\bea \label{SigmaMCh}
&&\Gamma^b_{\phi b} = \frac{1}{2} \quad , \quad \Gamma^b_{x b} = -1 \quad , \quad \Gamma^a_{g a} = - 1 \quad , \quad \Gamma^g_{aa} = e^{-2g} \nn \\
&&\Gamma^{\phi}_{bb} = - \frac{N_c^2 e^{\phi - 2 x}}{16} \quad , \quad \Gamma^x_{bb} = \frac{N_c^2 e^{\phi - 2x}}{32}
\eea
with all other components vanishing. Therefore, the only non-zero curvature components are:
\be \label{SigmaMC}
{\cal R}_{g a g a} = - \frac{e^{-2g}}{2} \,\, , \,\, {\cal R}_{\phi b \phi b} = - \frac{N_c^2 e^{\phi - 2x}}{128} \,\, , \,\, {\cal R}_{xbxb} = - \frac{N_c^2 e^{\phi - 2x}}{32} \,\, , \,\, {\cal R}_{xb \phi b} = \frac{N_c^2 e^{\phi - 2x}}{64} \,\, .
\ee
Now notice that all quantities in (\ref{EOM}), except for $(\delta \Phi)_{gi}^i$, are at zeroth order, i.e. are evaluated on the background. In other words, $\Phi'^k$ is non-vanishing only when the index $k$ corresponds to a scalar with a non-constant background dependence. So for us the only contribution of the term ${\cal R}^i{}_{kjl} \Phi'^k \Phi'^l$ is of the form ${\cal R}^i{}_{pjp} (\pd_z p_w)^2$, which due to (\ref{SigmaMC}) vanishes identically in our case. Similarly, the extended derivative $D_z$ simplifies to an ordinary partial derivative:
\be
D_z (\delta \Phi)_{gi}^i = \pd_z (\delta \Phi)^i_{gi} + \Gamma^i_{p j} (\pd_z p_w) (\delta \Phi)^j_{gi} = \pd_z (\delta \Phi)^i_{gi}
\ee
since there are no nonzero Christoffel symbol components with a $p$ index; see (\ref{SigmaMCh}). So the field equation for $i=g,x,\phi,a,b$ acquires the form:
\be \label{EM1}
\left[ \pd_z^2 + 4 \hat{A}' \pd_z + e^{-2\hat{A}} \pd_{\mu} \pd^{\mu} \right] \!(\delta \Phi)^i - V^i{}_{|j\neq p} (\delta \Phi)^j - \left[ V^i{}_{|p} + \frac{8}{\hat{A}'} V^i (\pd_z p_w) \right] (\delta p)_{gi} = 0 \,\, ,
\ee
whereas the field equation for $i=p$ becomes:
\bea \label{EM2}
&&\left[ \pd_z^2 + 4 \hat{A}' \pd_z + e^{-2\hat{A}} \pd_{\mu} \pd^{\mu} \right] \!(\delta p)_{gi} - \left[ V^p{}_{|j\neq p} + \frac{4 (\pd_z p_w) V_{j\neq p}}{3 \hat{A}'} \right] (\delta \Phi)^j - \nn \\
&&\left[ V^p{}_{|p} + \frac{8( \pd_z p_w ) V_p}{3\hat{A}'} + \frac{32 V (\pd_z p_w)^2}{3 \hat{A}'^2} \right] (\delta p)_{gi} = 0 \,\, .
\eea
Let us note again that all derivatives of the potential $V$ in (\ref{EM1})-(\ref{EM2}) have to be evaluated on the background values of the fields. In that regard, (\ref{Sol1})-(\ref{Sol2}) imply that on the walking background:
\be
\Gamma^g_{aa} = 1 \quad , \quad \Gamma^{\phi}_{bb} = - \frac{N_c^2}{A c^2} \quad , \quad \Gamma^x_{bb} = \frac{N_c^2}{2 A c^2} \quad .
\ee
Note that, although $\frac{N_c^2}{c^2} <\!\!< 1$, one has $A \sim c^{-3/4}$. So we cannot neglect $\Gamma^{\phi}_{bb}$ and $\Gamma^x_{bb}$.

Finally, before we begin solving equations (\ref{EM1})-(\ref{EM2}), we should perform in them the change of variables in (\ref{zdef}).

\section{Explicit linearized field equations}
\setcounter{equation}{0}

To write the field equations (\ref{EM1})-(\ref{EM2}) more explicitly, we need the derivatives of the potential (\ref{pot}). These are computed in Appendix \ref{A1}. Using (\ref{Vder})-(\ref{Vderder}), as well as (\ref{SigmaMCh}), we obtain the following equations of motion:
\bea
&&\left[ \pd_z^2 + 4 \hat{A}' \pd_z + e^{-2 \hat{A}} \pd_{\mu} \pd^{\mu} \right] \delta a - 2 V_{aa} \,\delta a - 2 V_{ab} \,\delta b \,= \,0 \quad , \nn \\
&&\left[ \pd_z^2 + 4 \hat{A}' \pd_z + e^{-2 \hat{A}} \pd_{\mu} \pd^{\mu} \right] \delta b - (G^{bb} V_{bb} + 2 V_{\phi} - V_x) \delta b - G^{bb} V_{ba} \,\delta a \,= \,0 \quad , \nn \\
&&\left[ \pd_z^2 + 4 \hat{A}' \pd_z + e^{-2 \hat{A}} \pd_{\mu} \pd^{\mu} \right] \delta g - 2 V_{gg} \,\delta g \,= \,0 \quad .
\eea

Similarly, for $\delta \phi$ we find:
\be
\left[ \pd_z^2 + 4 \hat{A}' \pd_z + e^{-2 \hat{A}} \pd_{\mu}\pd^{\mu} \right] \delta \phi - 4 V_{\phi \phi} \,\delta \phi - 4 V_{\phi x} \,\delta x - \left[ 4 V_{\phi p} + \frac{32 (\pd_z p_w)}{\hat{A}'} V_{\phi} \right] (\delta p)_{gi} = 0 \,\,\, .
\ee
One can easily verify that:
\be
\pd_z p_w = - \frac{2}{3} \frac{\pd \rho}{\pd z} \qquad {\rm and} \qquad \hat{A}' \equiv \pd_z \hat{A} = \frac{2}{3} \frac{\pd \rho}{\pd z} \quad ,
\ee
which implies:
\be
\frac{\pd_z p_w}{\hat{A}'} = - 1 \,\,\, .
\ee
Hence:
\be
4 V_{\phi p} + \frac{32 (\pd_z p_w)}{\hat{A}'} V_{\phi} = 0 \,\,\, .
\ee
So the field equation for $\delta \phi$ acquires the form:
\be
\left[ \pd_z^2 + 4 \hat{A}' \pd_z + e^{-2 \hat{A}} \pd_{\mu} \pd^{\mu} \right] \delta \phi - 4 V_{\phi \phi} \,\delta \phi - 4 V_{\phi x} \,\delta x = 0 \,\,\, .
\ee

Analogously, we have:
\be
\left[ \pd_z^2 + 4 \hat{A}' \pd_z + e^{-2 \hat{A}} \pd_{\mu} \pd^{\mu} \right] \delta x - V_{xx} \,\delta x - V_{x\phi} \,\delta \phi - \left( V_{xp} - 8 V_x \right) (\delta p)_{gi} = 0 \,\,\, ,
\ee
where
\be
V_{xp} - 8 V_x = 12 \left( e^{-4p_w -4x_w} - e^{2p_w - 2x_w} \right) \, .
\ee

Finally, the $(\delta p)_{gi}$ field equation is:
\bea
\left[ \pd_z^2 + 4 \hat{A}' \pd_z + e^{-2 \hat{A}} \pd_{\mu} \pd^{\mu} \right] (\delta p)_{gi} - \left( \frac{1}{6} V_{pp} - \frac{8}{3} V_p + \frac{32}{3} V \right) (\delta p)_{gi}&& \nn \\
- \left( \frac{1}{6} V_{p \phi} - \frac{4}{3} V_{\phi} \right) \delta \phi - \left( \frac{1}{6} V_{px} - \frac{4}{3} V_x \right) \delta x = 0&& \hspace*{-0.4cm}.
\eea
Using that
\be
V_{p \phi} = 8 V_{\phi} \,\,\, ,
\ee
we finally obtain:
\be
\left[ \pd_z^2 + 4 \hat{A}' \pd_z + e^{-2 \hat{A}} \pd_{\mu} \pd^{\mu} \right] (\delta p)_{gi} - \frac{1}{6} \left( V_{pp} - 16 V_p + 64 V \right) (\delta p)_{gi} - \frac{1}{6} \left( V_{px} - 8 V_x \right) \delta x = 0 \, .
\ee

Now let us perform the change of variables (\ref{zdef}) in all coefficients. First, note that:
\be
\hat{A}' = \frac{1}{4z} \, .
\ee
So the common differential operator for all field equations is:
\be
{\cal L}_z^2 \equiv \left[ \pd_z^2 + 4 \hat{A}' \pd_z + e^{-2 \hat{A}} \pd_{\mu} \pd^{\mu} \right] = \pd_z^2 + \frac{1}{z} \pd_z + \frac{1}{\sqrt{z}} \frac{2 \sqrt{6}}{A^{9/4} c^{5/4}}\, \pd_{\mu} \pd^{\mu} \, .
\ee
As for the terms arising from the potential, there are only four basic ingredients (as can be seen from Appendix \ref{A1}):
\bea
e^{-4p_w - 4x_w} &=& \frac{512}{3 (Ac)^{7/2}} \,z = \frac{512}{3^{15/8}} \left(\frac{\beta}{c}\right)^{7/8} z \, \equiv \, c_1 z \,\,\, , \nn \\
e^{2 p_w - 2x_w} &=& 16 \sqrt{\frac{3}{2}} \, \frac{1}{(Ac)^{9/4}} \, \frac{1}{\sqrt{z}} = \frac{16}{3^{1/16}\sqrt{2}} \left(\frac{\beta}{c}\right)^{9/16} \frac{1}{\sqrt{z}} \, \equiv \, \frac{c_2}{\sqrt{z}} \,\,\, , \nn \\
e^{8p_w - 2x_w + \phi_w} &=& \frac{36}{A^2 c^3} \,\frac{1}{z^2} = \frac{36}{\sqrt{3}} \frac{\sqrt{\beta}}{c^{3/2}} \,\frac{1}{z^2} \, \equiv \, \frac{16}{N_c^2} \, \frac{c_3}{z^2} \,\,\, , \nn \\
e^{8p_w} &=& \frac{9}{4} \,\frac{1}{Ac} \,\frac{1}{z^2} = \frac{9}{4 \times 3^{1/4}} \left( \frac{\beta}{c} \right)^{1/4} \frac{1}{z^2} \, \equiv \, \frac{c_4}{z^2} \,\,\, ,
\eea
where in the second equality we have used that $A\,c = \left(\frac{3\,c}{\beta}\right)^{1/4}$. Hence the six field equations become:
\bea
{\cal L}_z^2 \,\delta a &=& 2 \,c_3 \,z^{-2}\left( \delta a - \delta b \right) - \left( 2 \,c_2 \,z^{-1/2} - c_4 \,z^{-2} \right) \delta a \quad , \nn \\
{\cal L}_z^2 \,\delta b &=& - \,c_4 \,z^{-2}\left( 2 \,\delta a - \delta b \right) + \left( c_3 \,z^{-2} - 2 \,c_2\,z^{-1/2} + c_1\,z \right) \delta b \quad , \nn \\
{\cal L}_z^2 \,\delta g &=& 2 \left( c_1 \,z - c_2 \,z^{-1/2} + c_3 \,z^{-2} \right) \delta g \quad , \nn \\
{\cal L}_z^2 \,\delta \phi &=& - \,c_3 \,z^{-2} \left( 2 \,\delta x - \delta \phi \right) \quad , \nn \\
{\cal L}_z^2 \,\delta x &=& 4 \left( c_1 \,z - c_2 \,z^{-1/2} \right) \left[ \delta x - 3 \,(\delta p)_{gi} \right] + \frac{1}{2} \,c_3 \,z^{-2} \left( 2 \delta x - \delta \phi \right) \quad , \nn \\
{\cal L}_z^2 (\delta p)_{gi} &=& 2 \left( c_1 \,z - c_2 \,z^{-1/2} \right) \left[ \delta x + 3 \,(\delta p)_{gi} \right] \quad .
\eea

To restore all explicit $\beta$ dependence, let us now perform the change of variables $z \rightarrow u$, where $u$ was introduced in (\ref{Defu}). Defining the differential operator
\be
{\cal L}_u^2 = \pd_u^2 + \frac{1}{u} \pd_u + \frac{1}{u} \frac{c\, \beta}{48} \,\pd_{\mu} \pd^{\mu} \,\,\, ,
\ee
the equations of motion become:
\bea \label{lEoM2}
{\cal L}_u^2 \,\delta a &=& 2 \,\frac{f_3}{u^2} \left( \delta a - \delta b \right) - \left( 2 \,\frac{f_2}{u} - \frac{f_4}{u^2} \right) \delta a \quad , \nn \\
{\cal L}_u^2 \,\delta b &=& - \,\frac{f_4}{u^2} \left( 2 \,\delta a - \delta b \right) + \left( \frac{f_3}{u^2} - 2 \,\frac{f_2}{u} + f_1 \right) \delta b \quad , \nn \\
{\cal L}_u^2 \,\delta g &=& 2 \left( f_1 - \frac{f_2}{u} + \frac{f_3}{u^2} \right) \delta g \quad , \nn \\
{\cal L}_u^2 \,\delta \phi &=& - \,\frac{f_3}{u^2} \left( 2 \,\delta x - \delta \phi \right) \quad , \nn \\
{\cal L}_u^2 \,\delta x &=& 4 \left( f_1- \frac{f_2}{u} \right) \left[ \delta x - 3 \,(\delta p)_{gi} \right] + \frac{1}{2} \,\frac{f_3}{u^2} \left( 2 \delta x - \delta \phi \right) \quad , \nn \\
{\cal L}_u^2 (\delta p)_{gi} &=& 2 \left( f_1 - \frac{f_2}{u} \right) \left[ \delta x + 3 \,(\delta p)_{gi} \right] \quad ,
\eea
where
\bea\label{effs}
&&f_1 = \frac{3^{1/4} c^{1/4}}{144} \beta^{7/4} \quad , \quad f_2 = \frac{1}{12} \beta \, , \nn \\
&&f_3 = \frac{1}{\sqrt{3}} \,\frac{\sqrt{\beta}}{c^{3/2}} \,N_c^2 \quad , \quad f_4 = \frac{1}{ 3^{1/4}} \left( \frac{\beta}{c} \right)^{1/4} \, .
\eea

\section{Preliminaries on solving the field equations}
\setcounter{equation}{0}

Clearly, the field equations (\ref{lEoM2}) split into three separate groups. The equation for $\delta g$ is decoupled from the rest and can be solved analytically. The remaining equations are separated into two, independent of each other, coupled systems. One contains the pair $\delta a$, $\delta b$ and the other the fluctuations $\delta \phi$, $\delta x$ and $(\delta p)_{gi}$. The last two systems cannot be solved exactly. So, to make further progress analytically, we will use the same approximation techniques as in Section 4 of our paper \cite{ASW}. That is, we will solve exactly the approximate equations of motion for large and for small radial distance and match the two in an intermediate region, where they behave in the same way. The matching will produce the quantization condition for the masses of the various fields in the problem.

To implement the above procedure, we also need to impose two boundary conditions (for each field), in order to solve the second order differential equations. As in \cite{ASW}, we will impose that at large $\rho$, or in our context at the cut-off, the fields vanish since we are interested in perturbative states. The second boundary condition will be imposed at the lower end of the radial distance. To make a choice for it, let us first discuss our perspective. We are looking for a would-be dilaton in the spectrum of the walking region. In physical applications, different regions of the background geometry would correspond to different spectra, as going from one region to another would encode undergoing a phase transition in the dual field theory (for example, from extended technicolor to technicolor). So we will limit ourselves only to the walking metric (\ref{leadMetric}).\footnote{This is an essential difference compared to \cite{ENP,EP2}.} In effect, we will view it as an independent background, regardless of its origin, with a space-time radial coordinate $\rho$ running in $(0,\rho_{\Lambda})$.\footnote{Without the physical cutoff $\rho_{\Lambda}$ there would not be a discrete spectrum.} With that in mind, the natural boundary condition is the following. Since the lower end of the range of $\rho$ can be viewed as the origin in polar coordinates, the solution for any of the field functions $\delta \Phi \equiv \{ \delta a , \delta b , \delta g , \delta \phi , \delta x , (\delta p)_{gi} \}$ can only be smooth if $\frac{\pd \,(\delta \Phi)}{\pd \rho} = 0$ at $\rho=0$. This boundary condition was first pointed out in \cite{EW} and has since been widely used in the literature on glueball spectra computations from gravity duals. In terms of the variable $u = e^{4 \rho}$, introduced in (\ref{Defu}), the above Neumann boundary condition acquires the form:
\be \label{IRbc}
\frac{\pd (\delta \Phi)}{\pd u} \bigg|_{u=1} = \,0 \, .
\ee
Again, (\ref{IRbc}) has to be satisfied for each field, in order to have a smooth solution.

We should mention that there other options for boundary conditions. For example, the numerical work of \cite{ENP} uses Dirichlet boundary conditions $\delta \Phi \big|_{\rho =0} = 0$ for many of the field fluctuations, in order to ensure regularity. The later work \cite{EP} derived a more complicated set of boundary conditions (that are a mixture of Dirichlet and Neumann) by considering the effective five-dimensional description with an explicit IR cutoff $\rho_{IR}$, i.e. $\rho \ge \rho_{IR} >0$. This involves additional new terms living on the IR boundary for consistency. It should be noted, though, that the qualitative answer regarding the presence or absence of a light state is the same in both \cite{ENP} and \cite{EP}, despite the latter being more rigorous. So it is natural to expect that, in our case too, the simpler boundary conditions would not alter the qualitative behavior. This should not be surprising on physical grounds, as IR physics does not, normally, affect higher-energy spectra. Nevertheless, it is certainly worth exploring what the outcome would be if we imposed an IR cutoff and used the boundary conditions of \cite{EP}. We leave that for the future. Here we will only study (\ref{IRbc}), as well as $\delta \Phi \big|_{\rho =0} = 0$ for more generality.\footnote{Note that if we were to impose the boundary conditions of \cite{EP,EP2} (equation (22) of \cite{EP2}, to be more precise) at the UV end of our model, nothing would change compared to imposing just $\delta \Phi \big|_{\rho = \rho_{\Lambda}} = 0$. This will become more clear later on and we will comment on it at the appropriate place.} Finally, one should keep in mind that we are considering a supersymmetric configuration. In that regard, we have to make the same choice of boundary conditions (Dirichlet or Neumann) for all fields in each of the coupled systems.

\subsection{Large $u$ region}

To see what is the approximate form of the field equations for large $u$, let us compare the three kinds of terms $y_1 = f_1 $, $y_2 = f_2 \,u^{-1}$, $y_3 = f_3\, u^{-2}$ and $y_4 = f_4 \,u^{-2}$ at the cut-off and see which of them dominate(s). For this purpose, we will need to use that $\frac{N_c}{c} <\!\!< 1$ and $u_{\Lambda}^{-1} = {\cal O} (\beta)$, as well as $c>\!\!>1$ and $\beta <\!\!< 1$. However, these constraints alone do not lead to a well-defined behavior for all ratios $y_i / y_j$ with $i,j = 1,...,4$ , as we will see shortly. More precisely, in different regions of the parameter space $(c,\beta)$ different $y_i$ terms are dominant at large $u$.\footnote{In \cite{ASW}, it was thought that $c$ and $\beta$ have to satisfy the relation $c = \frac{\sqrt{3}}{16} \,\beta^{-1/2}$ in order to have axial-vector modes in the flavor sector. However, this conclusion resulted from a comparison between two Lagrangians without due care for the different overall normalizations. Doing the comparison carefully, one realizes that there is no need of a constraint and thus $c$ and $\beta$ remain independent parameters. We thank T. ter Veldhuis for bringing this issue to our attention.} Exploring the full parameter space at once is rather daunting analytically. So in this paper we will restrict ourselves to a particular case, in which things simplify. This will enable us to develop techniques and understanding that will be an invaluable guide in exploring the rest of the parameter space in the future.

The particular case we will explore here has the following natural motivation. Notice that all terms on the right-hand side of (\ref{lEoM2}) have a $u$-dependence of the form $u^0$, $u^{-1}$ or $u^{-2}$. Also, let us for the moment imagine that $u$ is a variable running in the interval $(0, \infty)$. Then, if the coefficients in front of the powers of $u$ are comparable to each other, the terms containing $u^{-2}$ would be negligible at large $u$, but dominant for $u \rightarrow 0$. In such a case, one could drop the $f_{3,4}$ terms at large $u$ and the $f_1$ terms at small $u$. The resulting system can be treated successfully by analytical means, which is our goal in order to achieve better understanding of the roles of the various fields. Of course, in our case things are somewhat more complicated since $u \in (1, u_{\Lambda})$ and also the coefficients in front of the various powers of $u$ contain the four quantities $f_i$. So, in order to reproduce the above situation of neglecting the $f_{3,4}$ terms for large $u$ and the $f_1$ terms for $u$ of order $1$, we will need to introduce certain restriction in the $(c,\beta)$ parameter space. More precisely, we will concentrate here on a particular case, that is characterized by a relation of the form $c \,\beta^p = {\cal O} (1)$ with some $p>0$. To see how this comes about, let us consider the behavior of the various ratios $y_i/y_j$ for $u$ of order $u_{\Lambda}$.

We begin with:
\be
\frac{y_4}{y_1}\bigg|_{u_{\Lambda}} = \frac{f_4}{f_1}\,u^{-2}_{\Lambda} \approx \beta^{1/2}\, c^{-1/2} <\!\!< 1
\ee
and
\be
\frac{y_4}{y_2}\bigg|_{u_{\Lambda}} = \frac{f_4}{f_2}\,u^{-1}_{\Lambda} \approx \beta^{1/4} \,c^{-1/4} <\!\!< 1 \, ,
\ee
which imply that in the large $u$ region the $y_4$ term is negligible compared to $y_{1,2}$\,, without any restriction on $c$ and $\beta$. Similarly, we have:
\be
\frac{y_2}{y_1}\bigg|_{u_{\Lambda}} = \frac{f_2}{f_1} \,u^{-1}_{\Lambda} \approx \beta^{1/4}\, c^{-1/2} <\!\! <1 \, .
\ee
Thus, the $y_2$ term can also be neglected compared to the $y_1$ term, if needed. Finally, the remaining two ratios are:
\be \label{y3y2}
\frac{y_3}{y_2}\bigg|_{u_{\Lambda}} = \frac{f_3}{f_2} \,u_{\Lambda}^{-1} \approx \frac{N_c^2}{c^{3/2}}\,\beta^{1/2} =\frac{N_c^2}{c^{2}}\,(\beta\, c)^{1/2}<\!\!<(\beta \, c)^{1/2}
\ee
and
\be
\frac{y_3}{y_1}\bigg|_{u_{\Lambda}} = \frac{f_3}{f_1} \,u^{-2} \approx \,\frac{N_c^2}{c^{2}}\,(\beta^3\, c)^{1/2}<\!\!<(\beta^3 \, c)^{1/2} \, .
\ee
For arbitrary $c>\!\!>1$ and $\beta <\!\!< 1$ these two ratios are clearly unconstrained. To ensure $y_3 <\!\!< y_{1,2}$\,, we need to impose a condition on the parameter space. It is easy to realize that the condition $c\, \beta \le {\cal O} (1)$ does the job for both ratios. Note also that with this condition all four $f_i$ in (\ref{effs}) are small quantities.

To summarize, at leading order of the small parameters $f_i$, we can neglect both $y_3$ and $y_4$ in the large $u$ region as long as the constraint $c\, \beta \le {\cal O} (1)$ is satisfied.

\subsection{Region $u\approx 1$}

Let us now perform a similar analysis for the region $u \approx 1$.
We find:
\bea \label{y1y3}
\frac{y_1}{y_2}\bigg|_{u=1} &=& \frac{f_1}{f_2} \approx \left(\beta^{3} \,c\right)^{1/4}, \nn \\
\frac{y_1}{y_3}\bigg|_{u=1} &=& \frac{f_1}{f_3} \approx \frac{1}{N_c^2} \,\beta^{5/4}\, c^{7/4} =\frac{1}{N_c^2} \left(c\,\beta^{5/7}\right)^{7/4}, \nn \\
\frac{y_1}{y_4}\bigg|_{u=1} &=& \frac{f_1}{f_4} \approx \beta^{3/2}\,c^{1/2} = \left(c\,\beta^3\right)^{1/2}\, .
\eea
Thus it is safe to neglect the $y_1$ term provided $c\,\beta^{5/7} \le {\cal O}(1)$. Note, in fact, that the last condition also implies $c \,\beta \le {\cal O} (1)$, which was the condition for neglecting $y_3$ at large $u$. We will see shortly that the $a-b$ system in (\ref{lEoM2}) can be easily solved analytically within the parameter range constrained by $c\,\beta^{5/7} \le {\cal O}(1)$, without neglecting $y_2$.

In addition, we also have the ratios:
\be \label{y2y4}
\frac{y_2}{y_4}\bigg|_{u=1} = \frac{f_2}{f_4} \approx \beta^{3/4}\,c^{1/4} =(c\,\beta^3)^{1/4}
\ee
and
\be \label{y2y3y4}
\frac{y_2}{y_3}\bigg|_{u=1} = \frac{f_2}{f_3} \approx \frac{1}{N_c^2}\,c^{3/2} \,\beta^{1/2}<\!\!< \left(c\,\beta^{1/3}\right)^{3/2} \, .
\ee
It is easy to realize that to ensure both inequalities $y_2 <\!\!< y_{3,4}$\,, it is sufficient to have $c \,\beta^{1/3} \le {\cal O} (1)$. This will be important for solving the $x-p-\phi$ system in (\ref{lEoM2}). In that regard, note that the condition $c \,\beta^{1/3} \le {\cal O} (1)$ implies also $c\,\beta^{5/7} \le {\cal O}(1)$. Therefore, both $y_1$ and $y_2$ are negligible within the parameter space constrained by $c \,\beta^{1/3} \le {\cal O} (1)$.

At this point, it is clear that the strongest condition $c \,\beta^{1/3} \le {\cal O} (1)$ ensures all needed inequalities. We want to note again, though, that the less restrictive condition $c\,\beta^{5/7} \le {\cal O}(1)$ will be enough for the $a-b$ system. Thus the latter will be solved in a slightly more general subspace of the full parameter space than the $x-p-\phi$ system.

\section{Solving the field equations}
\setcounter{equation}{0}

\subsection{The mass spectrum of $\delta g$}

To get an understanding of the nature of the solution, we start with solving the equation for $\delta g$, which does not mix with any of the other fluctuating fields. Thus, it is factorizable and solved by mass eigenstates. From (\ref{lEoM2}), we have:
\be\label{geq}
 \pd_u^2\delta g + \frac{1}{u} \pd_u\delta g + \frac{1}{u} \frac{c \beta}{48} \,m_g^2\,\delta g= 2 \left( f_1 - \frac{f_2}{u} + \frac{f_3}{u^2} \right) \delta g \, .
 \ee
  Let us examine this equation at large $u$, where it is dominated by the terms
\be
\pd_u^2\delta g=2\,f_1 \,\delta g \, .
\ee
The solution is:
\be
\delta g\simeq \alpha_1 \, e^{u\,\sqrt{2\,f_1}}+\alpha_2\,e^{-u\,\sqrt{2\,f_1}} \, .
\ee
Now, applying Dirichlet boundary conditions at the cutoff $u_{\Lambda}=e^{4\,\rho_\Lambda}$, we obtain:
\be\label{Dirichlet}
\alpha_1 \, e^{e^{4\,\rho_\Lambda}\,\sqrt{2\,f_1}}+\alpha_2\,e^{-e^{4\,\rho_\Lambda}\,\sqrt{2\,f_1}}=0.
\ee
The exponents can be evaluated at the cutoff as:
\be\label{value}
e^{4\,\rho_\Lambda}\,\sqrt{2\,f_1}\simeq\beta^{-1/8}\, c^{1/8}>\!\!>1 \, .
\ee
Consequently, in leading order the boundary condition is $\alpha_1=0$.  This is the same result we would get if the cutoff value of $\rho$ would be
$\rho_\Lambda=\infty$.\footnote{It is easily seen that the same result follows also from the Neumann condition $\partial_u (\delta g)|_{u=u_{\Lambda}} = 0$, where one can take $u_{\Lambda} \rightarrow \infty$. Furthermore, imposing the boundary condition (22) of \cite{EP2} in our case, leads simply to $\partial_u (\delta g)|_{u=u_{\Lambda}} = 0$, at leading order in $\beta$. Thus, for us the UV boundary condition does not play a crucial role.}

Then the principle for finding the mass eigenvalues is clear. Equation (\ref{geq}) has two independent solutions. One is exponentially increasing with $u$ and one is exponentially decreasing.  To satisfy the Dirichlet condition at infinity one must pick the exponentially decreasing solution.  Then we need to impose the Neumann or Dirichlet boundary conditions at $u=$1, providing the eigenvalue equation for the mass.

The solution of (\ref{geq}) can be given in terms of Whittaker functions,
\be
\delta g=\frac{1}{\sqrt{u}}\left[\alpha_1\,M_{\kappa,\mu}(u\,2\,\sqrt{2\,f_1})+\alpha_2\,W_{\kappa,\mu}(u\,2\,\sqrt{2\,f_1})\right],
\ee
where
\bea
\kappa&=&\frac{1}{2 \,\sqrt{2\,f_1}}\left(2\,f_2+\frac{c\,\beta\,m_g^2}{48}\right)\equiv \frac{8\,f_2+M^2}{8\,\sqrt{2\,f_1}}\nn\\
\mu&=&\sqrt{2f_3},
\eea
where we introduced $M^2=m^2\,c\,\beta\,/\,12$, a combination we will use in the paper along with $m^2$.  Loosely speaking we will also refer to $M$ as "mass."

The asymptotically decreasing solution is $W_{\kappa,\mu}(u\,2\,\sqrt{2\,f_1})\,/\,\sqrt{u}$.  Then the boundary condition, defining the mass spectrum for Neumann boundary conditions at $u=1$ is
\be\label{eigen}
\left.\frac{ d\, W_{\kappa,\mu}(u\,2\,\sqrt{2\,f_1})\, u^{-1/2}}{du}\right|_{u=1}=0
\ee
Since  $2\,\sqrt{2\,f_1}<\!\!<1$, we can expand in small argument the Whittaker function in (\ref{eigen}). To leading order and after omitting overall factors independent of $M$, we obtain:
\be\label{leadingN}\left.\frac{ d\, W_{\kappa,\mu}(u\,2\,\sqrt{2\,f_1})\, v^{-1/2}}{du}\right|_{u=1}\sim \frac{(8\,f_1)^{\mu/2}\,\Gamma(1-2\,\mu)}{\Gamma(1/2-\mu-\kappa)}+\frac{(8\,f_1)^{-\mu/2}\,\Gamma(1+2\,\mu)}{\Gamma(1/2+\mu-\kappa)}.
\ee

For Dirichlet boundary conditions at the IR end, the spectrum is also simple. Expanding the condition $W_{\kappa,\mu}(2\,\sqrt{2\,f_1})\,=0$ in $f_1$ we obtain:
\be\label{leadingD}
W_{\kappa,\mu}(2\,\sqrt{2\,f_1})\sim \frac{(8\,f_1)^{\mu/2}\,\Gamma(-2\,\mu)}{\Gamma(1/2-\mu-\kappa)}+\frac{(8\,f_1)^{-\mu/2}\,\Gamma(2\,\mu)}{\Gamma(1/2+\mu-\kappa)}.
\ee

Now, recall from (\ref{effs}) that 
\be 
\sqrt{f_3} \approx \frac{N_c}{c} \, (c \,\beta)^{1/4} <\!\!< (c \,\beta)^{1/4} \, . 
\ee
Therefore, as long as we are in the parameter space constrained by $c\,\beta^{1/3} \le {\cal O} (1)$ and thus also by $c\,\beta \le {\cal O} (1)$, we have $\sqrt{f_3} <\!\!< 1$. Then, due to $\mu=\sqrt{2\,f_3}<\!\!<1$, both terms in (\ref{leadingN}) and (\ref{leadingD}) change sign near
\be\label{spectrumg}
\frac{1}{2}+2\,n-\kappa \simeq\frac{1}{2}+2\,n-\frac{M^2+8\,f_2}{8\,\sqrt{2\,f_1}} \, .
\ee
This implies that the the $n$th mass level for both boundary conditions has the approximate form of
\be
M_n=m_n\,\frac{\beta^{1/2}\,c^{1/2}}{12}\simeq [2^{5/2}(1+2\,n)\,f_1^{1/2}-8\,f_2]^{1/2}.
\ee
Using (\ref{effs}) we obtain for the lowest mass
\be\label{deltagmass}
m_0\simeq \frac{2\,\sqrt{3}}{c^{1/2}} \,\beta^{-1/16}[2^{1/2}3^{1/8}\,c^{1/8}-4\,\beta^{1/8}]^{1/2}
\ee
It is interesting to compare (\ref{deltagmass}) with the mass of the vector boson mass in the flavor sector that we obtained in \cite{ASW},  $m_\rho \simeq {\text const } \times c^{-1/2}$.  The ratio $m_0\, /\,m_\rho$ rises very slowly with decreasing $\beta$.   We will find similar behavior for some of the masses of  low lying states in other sectors of fluctuations.

\subsection{The spectrum of the $\delta x$ - $\delta p$ - $\delta \phi$ system}

It is convenient to introduce the functions:
\be
\varphi_1 \equiv 2 \delta x - \delta \phi \qquad , \qquad \varphi_2 \equiv \delta x + 3 \delta p \qquad , \qquad \varphi_3 \equiv \delta x - 3 \delta p \quad .
\ee
In terms of these, the last three equations of (\ref{lEoM2}) can be written as:
\bea\label{phieqs}
{\cal L}_u^2 \,\varphi_1 &=& 8 \left( f_1 - \frac{f_2}{u} \right) \varphi_3 + 2 \,\frac{f_3}{u^2} \,\varphi_1 \, , \nn \\
{\cal L}_u^2 \,\varphi_2 &=& 4 \left( f_1 - \frac{f_2}{u} \right) \varphi_3 + 6 \left( f_1 - \frac{f_2}{u} \right) \varphi_2 + \frac{1}{2} \,\frac{f_3}{u^2} \,\varphi_1 \, , \nn \\
{\cal L}_u^2 \,\varphi_3 &=& 4 \left( f_1 - \frac{f_2}{u} \right) \varphi_3 - 6 \left( f_1 - \frac{f_2}{u} \right) \varphi_2 + \frac{1}{2} \,\frac{f_3}{u^2} \,\varphi_1 \, .
\eea

We separate the range of $u$, i.e. $1<u<u_{\Lambda}$\,, into high and low $u$ regions given by $u\geq u_\bullet$ and $1<u<u_\bullet$\,, respectively. In the former region we neglect $f_3$ and find the resulting solution $\varphi_i^h$, while in the latter region we drop $f_1,\,f_2$ and find $\varphi_i^l$. In Sec. 5 we showed that these approximations are valid as long as the condition $c\,\beta^{1/3} \le {\cal O}(1)$ is satisfied by the parameters $c$ and $\beta$.

Now, to match $\varphi_i^h$ with $\varphi_i^l$\,, we need to find an appropriate matching point. Natural choices are $u_1$ or $u_2$ where
\be
f_1\simeq \frac{f_3}{u_1^2},\quad  \frac{f_2}{u_2}\simeq \frac{f_3}{u_2^2} \, .
\ee
From (\ref{effs}), it follows that:
\bea
u_1 &\simeq &\frac{N_c}{c}\,\beta^{-11/16} \, ,\nn\\
u_2 &\simeq &\left(\frac{N_c}{c}\right)^2\,\beta^{-3/4} \, .
\eea
It is not certain whether $u_1>\!\!>1$ and $u_2>\!\!>1$. Clearly though, the exact choice of the matching point is not crucial, as long as the matched functions behave in the same way at $u_{\bullet}$. The matching process is facilitated by the observation that $\varphi_i^h$ are Whittaker functions of argument $\sim \sqrt{f_1}\,u$, while $\varphi_i^l$  are Bessel functions of argument $\sim m\,\sqrt{\beta \,c\,u}$.  Now it is easy to see that at $u=u_1$ or at $u=u_2$ these arguments   are much smaller  than one, provided $m_V\beta^{-1/2}>m$, where $m_V$ is the mass of the technirho.  Therefore, we can expand these functions to match $\varphi_i^h$ with $\varphi_i^l$.

\subsubsection{Solution for $1<u<u_\bullet$} \label{Low}

The equations at $u<u_\bullet$ take the form
\bea\label{low}
{\cal L}_u^2 \,\varphi_1^l &=&  2 \,\frac{f_3}{u^2} \,\varphi_1^l \, , \nn \\
{\cal L}_u^2 \,\varphi_2 ^l&=&  \frac{1}{2} \,\frac{f_3}{u^2} \,\varphi_1^l \, , \nn \\
{\cal L}_u^2 \,\varphi_3^l &=&  \frac{1}{2} \,\frac{f_3}{u^2} \,\varphi_1^l \, .
\eea
Solving the equation for $\varphi_1^l $ yields
\be\label{phi1l}
\varphi_1^l=c_1^J\,J_\nu(M\,\sqrt{u})+c_1^Y\,Y_\nu(M\,\sqrt{u}),
\ee
where
$\nu=2\,\sqrt{2\,f_3}$ and $M=m \,\sqrt{c\,\beta}\,/\,(2\,\sqrt{3}).$
Then using (\ref{phi1l}) we can solve (\ref{low}) for $\varphi_2^l$ and $\varphi_3^l$ to get
\bea\label{phi23l}
\varphi_2^l&=&\frac{1}{4}\varphi_1^l+c_2^J\,J_0(M\,\sqrt{u})+c_2^Y\,Y_0(M\,\sqrt{u}),\nn\\
\varphi_3^l&=&\frac{1}{4}\varphi_1^l+c_3^J\,J_0(M\,\sqrt{u})+c_3^Y\,Y_0(M\,\sqrt{u}).
\eea

\subsubsection{Solution for $u>u_\bullet$} \label{High}

The equations of motion simplify to
\bea\label{high}
{\cal L}_u^2 \,\varphi_1^h &=&  \left( f_1 - \frac{f_2}{u} \right) \,8\,\varphi_3 ^h , \nn \\
{\cal L}_u^2 \,\varphi_2^h &=&  \left( f_1 - \frac{f_2}{u} \right)\,(4\, \varphi_3^h + 6 \, \varphi_2 ^h), \nn \\
{\cal L}_u^2 \,\varphi_3^h &=&  \left( f_1 - \frac{f_2}{u} \right)\,(4\, \varphi_3^h - 6 \,\varphi_2^h) .
\eea
$\varphi_2^h$ and $\varphi_3^h$ can be found after  diagonalizing the last two equations. Combining fluctuations $\varphi_2^h$ and $\varphi_3^h$ into a two component vector, $\Phi$, we can write
\be\label{system23}
\Phi''+\frac{1}{u}\Phi'+\frac{M^2}{4\,u}\,\Phi=\left(f_1-\frac{f_2}{u}\right)\,{\cal M}\,\Phi,\nn\\
\ee
where
\be\label{matrix}
{\cal M}=\left(\begin{matrix} 4&6\\ 4&-6\end{matrix}\right),
\ee
Now the eigenvectors of $\cal M$ are
\be\label{combi}
\chi=\varphi_2^h+\alpha\,\varphi_3^h
\ee
and its complex conjugate $\chi^\star$, where, $\alpha$ and the eigenvalue corresponding to $\alpha$, $\lambda$, are
\be\label{alphalambda}
\alpha=\frac{1}{6}(1-i\,\sqrt{23}), {\text{ and }} \,\, \lambda=5+i\,\sqrt{23}.
\ee

The general diagonal solution is a combination of Whittaker functions, divided by $\sqrt{u}$, just like the solution for $\delta g$. Extending the upper limit to $\rho\to\infty$  the exponentially increasing solution can be discarded and we obtain the solution satisfying the correct Dirichlet boundary conditions at $u=\infty$ as
\be\label{hsolution}
\chi\sim \frac{e^{i\,\eta}}{\sqrt{u}}\,W_{\kappa,0}(2\,\sqrt{f_1\,\lambda}\,u),\nn\\
\ee
where
\be \label{kap}
\kappa=\frac{M^2+4\,\lambda\,f_2}{8\,\sqrt{\lambda\,f_1}}.
\ee
At this point the complex  phase, $\eta$, is arbitrary.

Using (\ref{combi}) and calculating $\varphi_2^h$ and $\varphi_3^h$ from  $\chi$ and $\chi^\star$ we obtain, using an arbitrary normalization, but fixed relative normalization
\bea\label{phi23h}
\varphi_2^h&=&i\frac{1}{\sqrt{u}}\,\left[\alpha^\star\,e^{i\,\eta}\,W_{\kappa,\,0}(2\,\sqrt{f_1\,\lambda}\,u)-\alpha\,e^{-i\,\eta}\,W_{\kappa^\star,\,0}(2\,\sqrt{f_1\,\lambda^\star}\,u)\,\right],\nn\\
\varphi_3^h&=&i\frac{1}{\sqrt{u}}\,\left[e^{-i\,\eta}\,W_{\kappa^\star,\,0}(2\,\sqrt{f_1\,\lambda^\star}\,u)-e^{i\,\eta}\,W_{\kappa,\,0}(2\,\sqrt{f_1\,\lambda}\,u),\right],
\eea

Next, adding the last two equations of (\ref{high}) we readily obtain the solution for $\varphi_1^h$,
\be\label{phi1h}
\varphi_1^h=\varphi_2^h+\varphi_3^h+d_1\left[J_0(M\,\sqrt{u})\,Y_0(M\,\sqrt{u_\Lambda})-Y_0(M\,\sqrt{u})\,J_0(M\,\sqrt{u_\Lambda})\right],
\ee
where $u_\Lambda=e^{4\,\rho_\Lambda}.$

\subsubsection{Matching method for the $\varphi$ system}

We will match $\varphi_i^l$ with $\varphi_i^h$ at an intermediate point $u_{\bullet}$, such that $1<\!\!<u_\bullet<\!\!<u_\Lambda$. Since we are looking for states with small mass, we will expand the solutions $\varphi$ in small $M$. This expansion in small argument converges very fast due to $\sqrt{f_1} \sim c^{1/8}\,\beta^{7/8} =\left(c\, \beta^7\right)^{1/8}\,<\!\!< 1$, since $c\,\beta^7<\!\!<c\,\beta^3={\cal O}(1)$ by our choice of the restricted space for the parameters $c$ and $\beta$. Also, the expanded form will allow us to perform the matching analytically.

All terms in $\varphi_i^{l,h}$, except for $\varphi_1^l$ in (\ref{phi1l}), have the following leading behavior: The two dominant terms are a constant (independent of $u$) and a term proportional to $\log (u)$.
However,  the leading terms of the expansion of $\varphi_1^l$  are proportional to non-integer powers of $u$. More specifically, recalling that the argument of the relevant functions is actually $M \sqrt{u}$, we have terms of the form  $(M^2 u)^{\pm\sqrt{2\,f_3}}$. Using
that $\sqrt{2\,f_3}<\!\!<1$, we can expand the powers of $u$ as
\be\label{phi1exp}
u^{\pm\sqrt{2\,f_3}}\simeq 1\pm{\sqrt{2\,f_3}}\,\log(u)+O[f_3\,\log^2(u)] \, ,
\ee
where terms of $O(f_3)$ are neglected, and the leading order terms of (\ref{phi1exp}) are $1$ and $\log(u)$. So $\varphi_1^l$ can also be matched smoothly to $\varphi_1^h$. An important remark is in order. The factors $M^{\pm2\,\sqrt{2f_3}}\equiv M^{\pm\nu}$ above should not be expanded in small $\sqrt{f_3}$. The reason is that for some choices of $N_c$ and $\beta$ the solution for $M$ may take values, such that $\log(M^{-1})\,\nu>1$, in which case an expansion of $M$ in $\sqrt{f_3}$ is not convergent. Therefore we keep the coefficients $M^{\pm\,\nu}$ intact during all subsequent calculations.

Before counting the number of constants and the number of equations, let us fix the boundary conditions at $u=1$.  No matter whether we choose Dirichlet or Neumann boundary conditions, the ratio of $c_1^J$ and $c_1^Y$ in (\ref{phi1l}) is fixed. In other words, $\varphi_1^l$ in (\ref{phi1l})  contains only one overall undetermined normalization constant, $c_1$. Then, turning to (\ref{phi23l}), we can see that imposing the same boundary conditions on $\varphi_{2}^l$ and $\varphi_{3}^l$ also requires fixing the ratios of $c_{2,3}^J\,/\,c_{2,3}^Y$. That leaves us with a single undetermined constant in each of $\varphi_2^l$ and $\varphi_3^l$, which we will denote by $c_2$ and $c_3$ respectively. So, in total, we have three constants in the low-$u$ region solutions.

Let us now look at the high-$u$ region. The expressions for $\varphi_2^h$ and $\varphi_3^h$ in (\ref{phi23h}) contain a single constant, the undetermined phase $\eta$. The most general solution is, in fact, given by multiplying both of those by a common overall constant. We have chosen to set the latter to one. This just amounts to fixing the overall normalization in all three matching equations. Finally, $\varphi_1^h$ in (\ref{phi1h}) has one additional constant, $d_1$. To recapitulate, the solutions in the two regions are characterized by five undetermined constants: $c_1$, $c_2$, $c_3$, $d_1$, and $\eta$.

Now, we must match each of the three $\varphi_i^h$ functions with the corresponding $\varphi_i^l$ function at $u=u_\bullet$.  As we discussed above, this produces six equations, three arising from the matching of the constant (i.e., independent of $u$) terms and the remaining three from the matching of the coefficients of the $\log (u)$ terms. Five equations can be used to eliminate the five normalization constants enumerated in the previous paragraph. Then, the remaining sixth equation depends only on the input parameters $\beta$, $N_c$ and $u_\Lambda$, as well as the mass parameter $M$. In other words, it provides an implicit equation for $M$ in terms of the input parameters. We denote this equation for Neumann and Dirichlet boundary conditions, respectively, as $F_{N,D}^\varphi(M,\beta, N_c)=0$.

\subsubsection{The derivation of the mass eigenvalue equations}

Now let us match the low and high $u$ solutions obtained in Sections \ref{Low} and \ref{High}. We will denote matching by the $\simeq$ sign. Also, for convenience, we denote $\nu=2\,\sqrt{2\,f_3}$. Then from (\ref{phi1l}), (\ref{phi23l}), (\ref{phi23h}) and (\ref{phi1h}) we have:
\bea\label{match1}
c_1\,L_\nu&\simeq &e^{i\,\eta}(\alpha^*-1) H_W+e^{-i\,\eta}\,(\alpha-1)\,H_W^*+d_1\,H_1\,,\nn\\
\frac{1}{4}c_1\,L_\nu+c_2\,L_0&\simeq&\alpha^*\,e^{i\,\eta}\,H_W+\alpha\, e^{-i\,\eta}\,H_W^*\,,\nn\\
\frac{1}{4}c_1\,L_\nu+c_3\,L_0&\simeq&- e^{i\,\eta}\,H_W- e^{-i\,\eta}\,H_W^*\,,
\eea
where we have introduced the following notation:
\bea\label{bits}
L_\nu&= &J_{\nu}(M\,\sqrt{u})-r_\nu\,Y_{\nu}(M\,\sqrt{u})\,,\nn\\
L_0&=& J_0(M\,\sqrt{u})-r_0\,Y_0(M\,\sqrt{u})\,,\nn\\
H_W&=&\frac{i}{\sqrt{u}}\,W_{\kappa,0}(2\sqrt{f_1\,\lambda}\,u)\,,\nn\\
H_1&=&J_0(M\,\sqrt{u})-r_H\,Y_0(M\,\sqrt{u})
\eea
with the coefficients $r_\nu$ and $r_0$ fixed to enforce Dirichlet or Neumann boundary conditions at $u=1$, while $r_H$ enforces Dirichlet boundary conditions at $u=u_\Lambda$. More precisely, we have:
\bea\label{ares}
r_\nu^D&=&\frac{J_{\nu}(M)}{Y_{\nu}(M)}\,,\nn\\
r_\nu^N&=&\frac{J_{-1+\nu}(M)-J_{1+\nu}(M)}{Y_{-1+\nu}(M)-Y_{1+\nu}(M)}\,,\nn\\
r_0^D&=&\frac{J_{0}(M)}{Y_{0}(M)}\,,\nn\\
r_0^N&=&\frac{J_{1}(M)}{Y_{1}(M)}\,,\nn\\
r_H&=&\frac{J_{0}(M\,\sqrt{u_\Lambda})}{Y_{0}(M\,\sqrt{u_\Lambda})}\,.
\eea
Note that the functions of (\ref{bits}) are completely fixed for each boundary condition at $u=1$. The unknown coefficients, $c_1$, $c_2$, $c_3$, $d_1$ and $e^{i\,\eta}$ are extracted from the bits that contribute to the matching equations and are explicitly displayed in (\ref{match1}). This will facilitate finding the mass eigenvalue equation in a symbolic form, before the explicit form of $L_\nu$, $L_0$, $H_1$ and $H_W$ is substituted.

One more operation is needed before we solve for the unknown coefficients. We need to expand the functions in (\ref{bits}) in small argument (due to $M <\!\!<1$) and pick out the constant and logarithmic terms. Using a collective notation $X$ for $L_0$, $L_\nu$, $H_1$ and $H_W$ we can write
\be\label{constpluslog}
X=X^0+X^1\,\log(u)+...
\ee
where the ellipsis denotes terms of higher order in $\log(u)$ and terms containing positive powers of $u$, all omitted in the matching procedure. Substituting  expansions (\ref{constpluslog}) into (\ref{match1}) we obtain six equations after separating the constant and $\log(u)$ terms. These are very similar to the three equations of (\ref{match1}) except we have superscripts 0 or 1 attached to $L_\nu$, $L_0$, $H_1$ and $H_W$.  Now we are ready to eliminate the 5 normalization constants and obtain a sixth equation, containing  a combination of $X^{0,1}$ but none of the five coefficients. For the details see Appendix \ref{MassCondPhi}. The final equation takes the form:
\bea\label{masseq}
&&0 = 4 ( \alpha^* - \alpha ) (L_0^1 H_W^{0*} - L_0^0 H_W^{1*}) (H_W^1 L_0^0 - H_W^0 L_0^1) (H_1^0 L_{\nu}^1 - H_1^1 L_{\nu}^0) \nn \\
&&- (1 + \alpha) (1 - \alpha^*) (L_0^0 H_W^{1*} - L_0^1 H_W^{0*}) (H_1^1 H_W^0 - H_1^0 H_W^1) (L_0^1 L_{\nu}^0 - L_0^0 L_{\nu}^1) \nn \\
&&+ (1 - \alpha) (1 + \alpha^*) (H_1^1 H_W^{0*} - H_1^0 H_W^{1*}) (H_W^1 L_0^0 - H_W^0 L_0^1) (L_0^1 L_{\nu}^0 - L_0^0 L_{\nu}^1) \, .
\eea

To analyze the last equation further, we need the expressions for the various components of $H$ and $L$. So let us write them down explicitly. Expanding the Bessel and Whittaker functions in (\ref{bits}) in small argument, we find:
\bea\label{components}
H_W^0&=&-i\frac{\sqrt{2}\,(f_1\,\lambda)^{1/4}}{\Gamma(1/2-\kappa)}\left[2\,\gamma+\log(2\,\sqrt{\lambda\,f_1})+\psi^0(1/2-\kappa)\right],\nn\\
H_W^1&=&-i\frac{\sqrt{2}\,(f_1\,\lambda)^{1/4}}{\Gamma(1/2-\kappa)}\,,\nn\\
H_1^0&=&1-\frac{2\,r_H}{\pi}\left[\gamma\,+\log(M/2)\right],\nn\\
H_1^1&=&-\frac{r_H}{\pi}\,,\nn\\
L_\nu^0&=&\left(\frac{M}{2}\right)^{\nu}\frac{1-r_\nu\,\cot(\nu\,\pi)}{\Gamma(1+\nu)}+\left(\frac{M}{2}\right)^{-\nu}\frac{r_\nu}{\pi}\Gamma(\nu)\,,\nn\\
L_\nu^1&=&-\left(\frac{M}{2}\right)^{-\nu}\frac{r_\nu\,\Gamma(1+\nu)}{2\,\pi}+\left(\frac{M}{2}\right)^{\nu}\frac{1-r_\nu\,\cot(\nu\,\pi)}{2\,\Gamma(\nu)}\,,\nn\\
L_0^0&=&1-\frac{2\,r_0}{\pi}\left[\gamma\,+\log(M/2)\right],\nn\\
L_0^1&=&-\frac{r_0}{\pi}\,,
\eea
where $\psi^{(0)}$ is the digamma function
\be
\psi^{(0)}(x)=\frac{\Gamma'(x)}{\Gamma(x)}\,.
\ee
Further, expanding $r_{\nu}$ and $r_0$ in small $M$, we have:
\bea\label{r1r2}
r_\nu^N&\simeq& \frac{M^{2\,\nu}\,\pi}{\Gamma(\nu)[M^{2\,\nu}\,\Gamma(1-\nu)\,\cos(\pi\,\nu)+4^\nu\,\Gamma(1+\nu)]}.\nn\\
r_0^N&\simeq&0,\nn\\
r_\nu^D&\simeq&\frac{M^{2\,\nu}\,\pi}{\Gamma(1+\nu)\,[M^{2\,\nu}\,\cos(\pi\,\nu)\Gamma(-\nu)+4^\nu\,\Gamma(\nu)]}.\nn\\
r_0^D&\simeq&\frac{\pi}{2[\gamma+\log(M\,/\,2)]}.
\eea
The above expansions are well-convergent for small $M$. However, the argument of the Bessel functions in $r_H$ is of order $1$ if $M$ is close to the masses of the flavor vector bosons. Therefore, we will keep the full expression for $r_H$ in (\ref{ares}).

Substituting (\ref{components}) and (\ref{r1r2}) in (\ref{masseq}), we obtain to leading order (and after quite some algebra):
\bea\label{final}
F_N^\varphi(M,\beta,\,N_c)&=&M^\nu\frac{8\,C\,r_H}{4^\nu\,\Gamma(1+\nu)+M^{2\,\nu}\,\cos(\pi\,\nu)\Gamma(1-\nu)}\,,\\
F_D^\varphi(M,\beta,\,N_c)&=&M^\nu \frac{ C\,B\,B^*\,\nu\,\left\{\pi-2\,r_H[\gamma+\log(M\,/\,2)]\right\}}{[4^\nu\,\Gamma(1+\nu)-M^{2\,\nu}\,\cos(\pi\,\nu)\Gamma(1-\nu)][\gamma+\log(M\,/\,2)]^2}\,,\nn
\eea
where we have defined
\bea
C&=&\frac{2^{\nu+1}\sqrt{2\,f_1}(\alpha-\alpha^*)\,\lambda^{1/4}\,\lambda^{*\,1/4}}{\pi\,\Gamma(1/2-\kappa)\,\gamma(1/2-\kappa^*)}\,,\nn\\
B&=&2\,\gamma +\log(2\,\sqrt{2\,f_1\,\lambda})+\psi^{(0)}(1/2-\kappa)\,.
\eea

Note that to obtain the above equations we had to divide by the multiplier $(L_0^1 L_{\nu}^0 - L_0^0 L_{\nu}^1)$; see Appendix \ref{MassCondPhi}. The important point about this observation is that due to (\ref{components}) and (\ref{r1r2}) this multiplier is proportional to $M^{\nu}$. Therefore, the expressions  in (\ref{final}) have been derived under the assumption that $M^{\nu}\neq 0$. In other words, there is no $M=0$ solution of the constraints $F_{N,D}^\varphi(M,\beta,\,N_c)=0$, as they are not valid for that value of the parameter $M$.

The constants $C$ and $B$ have no zeros and the denominators do not have poles for {\em real} $M$.  However, for Neumann boundary conditions there are additional solutions at $r_H\sim J_0(M\,\sqrt{u_\Lambda})=0$, which is exactly the equation for the technirho and its KK excitations. Considering that $u_\Lambda= u_\star \,c_ \Lambda \equiv 2\,c_\Lambda \,/\,\beta$, where $0.01<c_\Lambda<1$ defines the end of the walking region \cite{ASW}, the lowest mass in this sector is given by,
\be\label{constantsolution}
m_0\simeq (\beta\, c)^{-1/2}\, r_1^J\, u_\Lambda^{-1/2}\sim c^{-1/2}.
\ee
where $r_1^J$ is the first zero of $J_0(x)$. This is the second type of behavior for the mass spectrum, after (\ref{deltagmass}) that we found for the system of fluctuations.  The $\beta$-dependence of this solution is the same as most of the solutions found in \cite{ENP} by numerical methods.

 Furthermore, there are additional solutions for Dirichlet boundary conditions, as well, at
\be\label{dir}
\frac{2}{\pi} J_0(M\,\sqrt{u_\Lambda})= \frac{Y_0(M\,\sqrt{u_\Lambda})}{\gamma+\log(M\,/\,2)}.
\ee
This equation agrees with the equation for  the axial vector states of the flavor sector.  When $\beta\to0$ $M=m\,\sqrt{\beta\,c\,/\,12}$ also decreases, consequently the denominator of the right hand side of (\ref{dir}) becomes a large negative number, driving the the argument  of the Bessel function on the left hand side towards the zero from above.  Consequently, the solution of (\ref{dir}) coincides with (\ref{constantsolution}) asymptotically.

\subsection{The spectrum of the $\delta a$ - $\delta b$ system}

Let us now turn to the first two equations in (\ref{lEoM2}). As for the $\varphi$-system, we will solve them separately in the low $u$ (IR) and the high $u$ (UV) regions and then match the two solutions. However, since $f_2$ enters both equations in the same way, it can be absorbed in the definition of the differential operator $\cal L$ by the substitution $M^2\to \hat M^2=M^2+8 \,f_2$, where $M=m\,\sqrt{c\,\beta\,/\,12}$ as before. Then all of the equations, describing the $\delta a - \delta b$ system, become dependent only on the combination $\hat M$, and not $M$ or $f_2$ separately.

Now, having only $f_1$, $f_3$ and $f_4$ terms on the right-hand side, we separate the low and high $u$ regions similarly to the $\varphi$ system. Namely, in the UV we keep only the $f_1$ term, whereas in the IR we keep only the $f_3$ and $f_4$ terms. This allows us to solve the equations analytically. The strongest restriction of the parameter space, that ensures the above approximations, is implied by the condition $f_1\,/\,f_3<\!\!<1$ at low $u$. As we saw in Section 5, this means that we restrict our considerations to the part of parameter space constrained by $\beta^5\,c^7={\cal O}(1)$. Note, again, that this is somewhat more general than the condition $c\,\beta^{1/3} = {\cal O} (1)$, relevant for the $\varphi$ system.

At high $u$ (i.e., in the UV) we find the following general solution:
\bea\label{deltaabh}
(\delta a)_h &=& C_a \,H_a\equiv C_a\,\left[\,J_0 \left( \hat M \,\sqrt{u} \right) -r_H ^a\,Y_0 \left(  \hat M\,\sqrt{u} \right)\right], \nn \\
(\delta b)_h &=& C_b\,H_b\equiv C_b\,\frac{1}{\sqrt{u}}\left[\,W_{\tilde{\kappa}, 0} \left( 2 \sqrt{f_1} u \right) -r^b_H \,M_{\tilde{\kappa},0} \left( 2 \sqrt{f_1} u \right) \,\right] ,
\eea
where
\be
\tilde{\kappa} = \frac{\hat M^2 }{8 \sqrt{f_1}} \,\, .
\ee
We choose the coefficient $r_H^a$ such that the boundary condition $(\delta a)_h = 0$ is satisfied at $u=u_{\Lambda}$:
\be
r_H^a=\frac{J_0 \left( \!\hat M \,\sqrt{u_\Lambda} \right)}{Y_0 \left( \!\hat M\,\sqrt{u_\Lambda} \right)}\,.
\ee
In addition, we take $r^b_H=0$, as for the $\varphi$-system, in order to eliminate the exponentially increasing term from $(\delta b)_h.$ Finally, we can choose $C_b=1$ to set the overall normalization scale.

The low $u$ solution is:
\bea\label{deltaal}
(\delta a)_l &=&C_1\,L_{\nu_1}+C_2\,L_{\nu_2}\equiv C_1\,\left[ \,J_{\nu_1}\left( \!\hat M \,\sqrt{u} \right) - r_{\nu_1}\,Y_{\nu_1}\left( \! \hat M\,\sqrt{u} \right)\right]\nn\\
& +&C_2\left[\,J_{\nu_2} \!\left(\!\hat M \,\sqrt{u} \right) -r_{\nu_2}\,Y_{\nu_2} \!\left( \!\hat M\,\sqrt{u} \right) \right]\, ,
\eea
where
\be
\nu_1 = \sqrt{6 f_3 + 4 f_4 - 2 \sqrt{f_3 (16 f_4 + f_3)}} \quad , \quad \nu_2 = \sqrt{6 f_3 + 4 f_4 + 2 \sqrt{f_3 (16 f_4 + f_3)}} \,\, ,
\ee
and
\be\label{deltabl}
(\delta b)_l =\!\frac{1}{4} \left( 1+ \sqrt{\frac{16 f_4 + f_3}{f_3}} \,\right)\,L_{\nu_1}+\frac{1}{4} \left( 1- \sqrt{\frac{16 f_4 + f_3}{f_3}} \right)\,L_{\nu_2}\,,
\ee
where $L_{\nu_1}$ and $L_{\nu_2}$ were introduced in (\ref{deltaal}). In the above, $r_{\nu_1}$ and $r_{\nu_2}$ are determined by imposing Dirichlet or Neumann boundary conditions at $u=1$. The constants  $r_{\nu_1}$ and $r_{\nu_2}$ can be obtained from (\ref{ares}) with the substitutions $\nu\to\nu_1$ and $\nu\to\nu_2$, respectively.

Now let us match the high and low $u$ solutions at a point $1<\!\!<u_\bullet<\!\!<u_\Lambda$. We obtain the following conditions:
\bea\label{matchingab}
C_a\,H_a&\simeq&C_1\,L_{\nu_1}+C_2\,L_{\nu_2}\,,\nn\\
H_b&\simeq& \,C_1\,F_1\,L_{\nu_1}+C_2\,F_2 \,L_{\nu_2}\,,
\eea
where for convenience we introduced the notation:
\be
F_1=\left( 1+ \sqrt{\frac{16 f_4 + f_3}{f_3}}\,\right) \quad , \quad F_2=\left( 1- \sqrt{\frac{16 f_4 + f_3}{f_3}}\,\right).
\ee

As for the $\varphi$-system, we will expand each of $H_a$, $H_b$, $L_{\nu_1}$ and $L_{\nu_2}$ for small argument. The leading terms are again constant and logarithmic ones. Equating the corresponding expansion terms on both sides of (\ref{matchingab}), we obtain four conditions. Three of them can be used to eliminate the constants $C_1$, $C_2$ and $C_a$. Then the fourth condition gives
the following mass eigenvalue equation:
\be\label{massab}
0=F_1\,(H_a^0\,L_{\nu_2}^1-H_a^1\,L_{\nu_2}^0)\,(H_b^0\,L_{\nu_1}^1-H_b^1L_{\nu_1}^0)+F_2(H_a^1\,L_{\nu_1}^0-H_a^0\,L_{\nu_1}^1)\,(H_b^0\,L_2^1-H_b^1L_{\nu_2}^0)\,.
\ee
Here, as in (\ref{constpluslog}), an upper index``$0$" denotes the constant term in the expansion of the corresponding quantity, whereas an upper index ``$1$" denotes the coefficient of the $\log u$ term.

To obtain the specific form of (\ref{massab}) for Dirichlet and Neumann boundary conditions, let us expand all of $r_{\nu_1}$,  $L_{\nu_1}^0$, $L_{\nu_1}^1$,  $r_{\nu_2}$,  $L_{\nu_2}^0$, $L_{\nu_2}^1$ as in (\ref{ares}) and (\ref{components}).
Keeping only the leading terms in powers of $\hat M$ and only multipliers which depend on $\hat M$, we find for Neumann boundary conditions:
\be\label{abN}
F_N^{ab}\sim r_H^a\frac{\hat M^{\nu_1+\nu_2}}{\Gamma(1/2-\tilde\kappa)\,g_N(\nu_1)\,g_N(\nu_2)}=0\,,
\ee
where
\be
g_N(\nu)=\hat M^{2\,\nu}\,\cos(\pi\,\nu)\,\Gamma(1-\nu)+4^{\nu}\Gamma(1+\nu)
\ee
and we have omitted a constant prefactor.

Similarly, for Dirichlet boundary conditions we obtain:
\be\label{abD}
F_D^{ab}=\frac{\hat M^{\nu_1+\nu_2}\,h_1(\hat M)\,h_2(\hat M)}{\Gamma(1/2-\tilde{\kappa})\,g_D(\nu_1)\,g_D(\nu_2)}=0\,,
\ee
where
\bea \label{ghh}
g_D(\nu)&=&\hat M^{2\,\nu}\,\cos(\pi\,\nu)\,\Gamma(-\nu)+4^{\nu}\Gamma(\nu)\,,\nn\\
h_1(\hat M)&=&\frac{2}{\pi}\,r_H^a\left[\gamma+\log\left(\frac{\hat M}{2}\right)\right]-1\,,\nn\\
h_2(\hat M)&=&2\,\gamma+\log(2\,\sqrt{f_1})+\psi^{(0)}(1/2-\tilde{\kappa})\,.
\eea

Note that, as in the case of the $\varphi$-system, the intermediate computations that lead to (\ref{abN}) and (\ref{abD}) involve division by $\hat M^{\nu_1+\nu_2}$. So the constraints $F_N^{ab} = 0$ and $F_D^{ab} = 0$ are only valid for $\hat M^2 \neq 0$. In fact, the matching conditions do not have a solution for $\hat M^2=0$.

The mass spectra for Neumann and Dirichlet boundary conditions can be read off from (\ref{abN}) and (\ref{abD}) respectively. Let us consider each case in turn. We begin with Neumann boundary conditions. The solutions of (\ref{abN}) are given by the zeros of $r_H^a$ and by the poles of $\Gamma(1/2-\tilde \kappa)$. The spectrum due to $r_H^a = 0$ is the same as the one for the vector mesons of the flavor sector, found in \cite{ASW}. A potential conceptual difference might have been expected due to the fact that the mass parameter $M$ now enters only through the combination $\hat{M}^2 = M^2 + 8 f_2$. So it is conceivable to have solutions with $\hat{M} < 8 f_2$ and thus $M^2 < 0$. However, that does not happen within the range of validity of our approximations. Indeed, the smallest value of $M^2$ that solves $r_H^a = 0$ is:
\be\label{critical}
M^2= \frac{(c_J^0)^2}{u_\Lambda}-8\,f_2=\beta\left(\frac{(c_J^0)^2}{2\,c_\Lambda}-\frac{2}{3}\right),
\ee
where $c_J^0$ is the smallest zero of the Bessel function $J_0$ and we have introduced the notation
\be
c_\Lambda= \frac{u_\Lambda}{u_*}=e^{4(\rho_\Lambda-\rho_*)}
\ee
with $\rho_* = \frac{1}{4} \log (2 \beta)$ as in \cite{ASW}. Note that within our approximations one has $c_{\Lambda} <\!\!< 1$, see \cite{ASW}. On the other hand, it is easy to convince oneself that (\ref{critical}) implies $M^2>0$ for every $c_\Lambda < 3 \, (c_J^0)^2\,/\,4\simeq 4.3.$ Therefore, the spectrum arising from the solutions of $r_H^a = 0$ does not contain a tachyonic mode.

\begin{figure}[htbp]
\begin{center}
\includegraphics[width=4in]{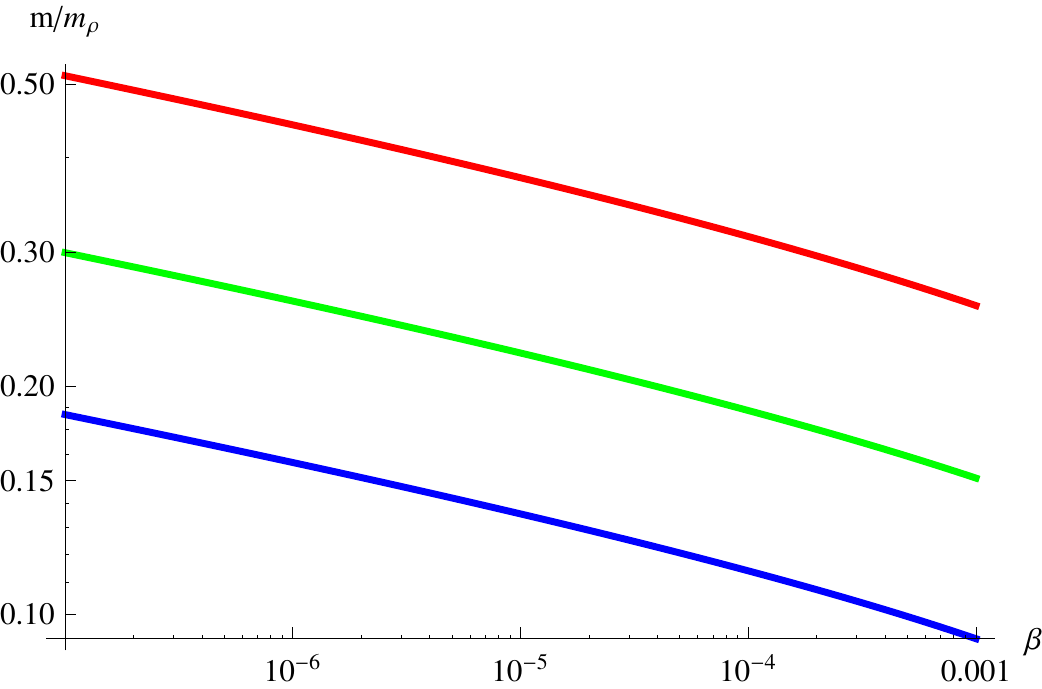}
\caption{The ratio of the mass of the lightest glueball in the $\delta a-\delta b$ sector with Neumann boundary conditions to the mass of the technirho, as a functions of $\beta$ at three different choices  of the parameter $c_\Lambda$, 0.02 (blue curve), 0.06 (green curve) and 0.2 (red curve).}
\label{kappa}
\end{center}
\end{figure}

The second series of states for Neumann boundary conditions is obtained from the solutions of the equation $\tilde \kappa = 1/2+n$ with $n$ integer, as mentioned above. These are slightly lighter than the corresponding states in the vector-meson spectrum. For illustration, we plot the ratio of the lowest mass in the $\tilde \kappa = 1/2+n$ spectrum, obtained for $\tilde\kappa=1/2$, to the mass of the lightest flavor vector-meson as a function of $\beta$ in Fig. \ref{kappa}. In this plot, we have fixed the parameter $c$ in the following way. We chose the value of $c$ such that the electroweak $S$-parameter, given in \cite{ASW} as
\be
S\approx \frac{\pi}{2\,(2\,\pi)^{3}\,3^{3/4}} c^{5/4}\,\beta^{-1/4}\,c_\Lambda\frac{1}{\log(u_\Lambda)},
\ee
is equal to the maximal phenomenologically admissible value, i.e. $S=0.09$. It is easy to verify that this choice of $c$ satisfies all of the constraints in Sec. 5, that are required for the consistency of our method of solving the equations of motion. Lowering the value of $S$ decreases  slightly the ratio plotted in Fig.\ref{kappa}.
Note that, again, within the range of validity of our approximations, the spectrum arising from $\tilde{\kappa} = 1/2$ does not contain a tachyon. Indeed, the mass of the lightest state is:
\be\label{mabN}
M_N^2=4\,\sqrt{f_1}-8\,f_2=\frac{\beta}{3}\left[\left(\frac{3\,c}{\beta}\right)^{1/8}-2\right].
\ee
So, to have $M_N^2 < 0$, one needs $\beta> \frac{3c}{2^8}$.  Recall, however, that the model we study \cite{NPP} is valid only for $c >\!\!> 1$ and we choose a large walking region by assuming $\beta<\!\!<1$. Then (\ref{mabN}) implies that there are no parametrically zero mass or tachyon solutions in the system.

There is a hint that zero mass states may appear if the range of applicability of our approximations were extended. Namely, notice that if we used (\ref{critical}) and (\ref{mabN}) beyond the range of their applicability, say at $c_\Lambda=O(1)$ or at $c/\beta=O(1)$, then zero mass states may appear in the spectrum. This might indicate that the appearance of a parametrically zero mass state in \cite{ENP} is related to the absence of a cutoff in the radial variable or the range of the parameter space used.

Now we turn to the spectra in the case of Dirichlet boundary conditions. These arise from the solutions of $h_1 (\hat M) = 0$ and $h_2 (\hat M) = 0$, as can be seen from (\ref{abD}). Note that now there are no solutions from the poles of $\Gamma(1/2-\tilde \kappa)$ in the denominator as these are canceled by the poles of $\psi^{(0)}(1/2-\tilde{\kappa})$ within $h_2(\hat M)$ of the numerator. The spectrum due to $h_1 = 0$ is the same as the axial-vector meson spectrum of the flavor sector, obtained in \cite{ASW}. Let us now take a more careful look at the spectrum due to the solutions of $h_2 = 0$. For illustrative purposes, let us focus on the lightest state. Since for small $\beta$ the term $\log(\sqrt{f_1})$ is large and negative, the digamma function must be close to its first pole in order to have $h_2 (\hat{M})$ of (\ref{ghh}) vanish. Thus, to leading order in a small argument expansion, we can substitute $\psi^{(0)}(1/2-\tilde \kappa)\simeq -\gamma-1/(\frac{1}{2}-\tilde \kappa)$. This gives the following approximate solution for $M_D^2$:
\be
M_D^2\simeq 4\sqrt{f_1}\left(1-\frac{2}{\gamma+\log(2\,\sqrt{f_1})}\right)-8\,f_2.
\ee
As before, for the physically acceptable range of values of our parameters (i.e., within the bounds of our approximations), the last expression is always positive. Note also that the only difference with the case of Neumann boundary conditions is the term $-\frac{2}{\gamma+\log(2\,\sqrt{f_1})}>0$ in the brackets. Hence the states for Dirichlet boundary conditions are somewhat heavier than those for Neumann ones.

\subsection{Summary of the mass-spectra of fluctuations}

We have investigated the mass spectra in the three independent sectors of fluctuations, $\delta g$,  the $\delta x - \delta p - \delta \phi$-system and the $\delta a - \delta b$-system.  We have found two different types of solutions.  The first of these turned out to be very similar to the solutions of the flavor sector that we investigated earlier \cite{ASW}. These solutions have a mass spectrum
\be \label{Msp}
m_n\sim a_n\,c^{-1/2},
\ee
where $a_n$ are constant. The second type of mass spectra are of the form:
\be
m_n\sim a_n\,\beta^{-1/16}\,c^{-7/16}.
\ee

Finally, it should be noted that under the conditions we use (cutoff in the radial coordinate and restrictions of the parameters space) we do not find a state that is parametrically lighter than the rest of the spectrum and thus could be viewed as a candidate dilaton.

\section{On the existence of a dilaton} \label{ED}
\setcounter{equation}{0}

For completeness, let us now see whether there are fluctuations that couple to the color field strength like a dilaton would. To do that, we will consider the embedding of a susy D5 probe into the fluctuated 10d background.

First, recall that the 10d metric is:
\bea
ds^2 &=& e^{2p -x} g_{IJ} dx^I dx^J + e^{x+g} (d\theta^2 + \sin^2 \theta d\varphi^2) + e^{x-g} \left[ \left( \tilde{\omega}_1 + a d\theta \right)^2 + \left( \tilde{\omega}_2 - a \sin \theta d \varphi \right)^2 \right] \nn \\
&+& e^{-6p-x} \left( \tilde{\omega}_3 + \cos \theta d\varphi \right)^2 \,\, ,
\eea
where, expanding around the walking background, we have:
\be \label{Fluct}
p = p_w + \delta p \quad , \quad x = x_w + \delta x \quad , \quad g = \delta g \quad , \quad a = \delta a \quad .
\ee
In addition, the expansion of the 5d metric is the following:
\be
g_{IJ} dx^I dx^J = N^2 dz^2 + \tilde{g}_{\mu \nu} (dx^{\mu} + N^{\mu} dz) (dx^{\nu} + N^{\nu} dz) \quad ,
\ee
where
\be \label{Metric1stOr}
N = 1 + n \quad , \quad N^{\mu} = n^{\mu} \quad , \quad \tilde{g}_{\mu \nu} = e^{2 \hat{A}} \left[ (1+\zeta) \eta_{\mu \nu} + \gamma_{\mu \nu} \right]
\ee
with $n$, $n^{\mu}$, $\zeta$ and $\gamma_{\mu \nu}$ being first order quantities. Also, $\gamma_{\mu \nu}$ is traceless.

Now, the supersymmetric D5 probe worldvolume is characterized by (\ref{sigma2}), as well as $\rho = const$ and thus $z = const$. Hence, the induced metric on the its worldvolume is:
\bea \label{D5ind}
ds^{2,\,induced}_{D5} &=& e^{2p -x} \tilde{g}_{\mu \nu} dx^{\mu} dx^{\nu} + \left[ e^{x+g} + e^{x-g} (a-1)^2 \right] (d\theta^2 + \sin^2 \theta d\varphi^2 ) \\
&=& e^{2p - x + 2\hat{A}} (1 + \zeta) \eta_{\mu \nu} dx^{\mu} dx^{\nu} + \left[ e^{x+g} + e^{x-g} (a-1)^2 \right] (d\theta^2 + \sin^2 \theta d\varphi^2 ) \,\, , \nn
\eea
where we have turned off the fluctuation $\gamma_{\mu \nu}$, since we are looking for a coupling of the form $ {\cal D} F_{\mu \nu} F^{\mu \nu}$ with ${\cal D}$ the techni-dilaton and $F^{\mu \nu} = \eta^{\mu \tilde{\mu}} \eta^{\nu \tilde{\nu}} F_{\tilde{\mu} \tilde{\nu}}$. The DBI action for the probe D5 brane is:
\be \label{D5DBI}
S_{D5}^{DBI} = - T_{D5} \int d^4 x d\Sigma_2 \,e^{-\phi} \sqrt{-\det (g^{induced}_{ab} + (2 \pi \alpha') F_{ab})} \,\, .
\ee
Expanding to second order in $F_{\mu \nu}$, we have:
\be
e^{-\phi} \sqrt{-\det (g^{ind}_{ab}+ (2 \pi \alpha' ) F_{ab})} = e^{-\phi} \sqrt{-\det g_{ab}^{ind}} \left[ 1 - \frac{1}{4} (2\pi \alpha')^2 g_{ind}^{\mu \tilde{\mu}} \,g_{ind}^{\nu \tilde{\nu}} \,F_{\mu \nu} F_{\tilde{\mu} \tilde{\nu}} \right] .
\ee
Using (\ref{D5ind}) and substituting (\ref{Fluct}), as well as $\phi = \phi_w + \delta \phi$, we find to first order:
\bea
e^{-\phi} &=& e^{-\phi_w} (1 - \delta \phi) \,\,\, , \nn \\
\sqrt{-\det (g^{ind}_{\mu \nu})} &=& e^{2 \phi_w} (1 + 4 \delta p - 2 \delta x + 2 \zeta) \,\,\, , \nn \\
\sqrt{\det (g^{ind}_{\theta \varphi})} &=& 2 e^{x_w} (1 + \delta x - \delta a) \, \sin \theta \,\,\, , \nn \\
g_{ind}^{\mu \nu} &=& e^{-\phi_w} (1 - 2 \delta p + \delta x - \zeta) \eta^{\mu \nu} \,\,\, .
\eea
Collecting everything together, we obtain from the action (\ref{D5DBI}) the following coupling (to first order) between the background fluctuations and $F^2$:
\be
{\cal L} \,\supset \,const \, e^{-\phi_w + x_w} (\delta x - \delta a - \delta \phi) F_{\mu \nu} F^{\mu \nu} \,\,\, .
\ee
In other words, we find a coupling of the form:
\be \label{DFF}
{\cal D} F_{\mu \nu} F^{\mu \nu} \,\,\, ,
\ee
where we have introduced the scalar
\be \label{TD}
{\cal D} = \delta x - \delta a - \delta \phi \, .
\ee
The coupling (\ref{DFF}) is of the form that is expected for the dilaton. Note that the fluctuations $\delta g$, $\delta p$ and $\zeta$ dropped out. This is especially nice for $\delta p$ and $\zeta$, since these are the only gauge-variant scalars in our case. The gauge-invariant quantity $(\delta p)_{gi}$ is a combination of them.

From (\ref{DFF})-(\ref{TD}) one can conclude that a prospective dilaton in this model is a combination of states from the $\delta a$-$\delta b$ and $\delta \varphi$ sectors. However, this combination is not a mass eigenstate. It could be that at low energies the wavefunction of ${\cal D}$ is dominated by one of the two sectors and thus corresponds to a state with an approximately well-defined mass. Even if that were the case, though, we have already seen that, within the parts of parameter space that were explored here, neither of those sectors has a light scalar
that could be identified with a slight breaking of conformal invariance.

\section{Discussion}

We investigated analytically the scalar glueball spectrum of the background of \cite{NPP}, corresponding to a walking gauge theory. Using the same approximations as in \cite{ASW,ASW2}, we found that the initial system of six coupled equations actually splits into three independent subsystems. To make further analytic progress, which was our goal in order to achieve better understanding, we had to concentrate on a subspace of the full parameter space of our backgrounds. This corresponds to a restriction on the constants $c$ and $\beta=\sin^3\alpha$, namely a bound of the form $c\,\beta^{1/3} \le {\cal O}(1)$.\footnote{As we stressed previously, for the $a-b$ sector the less restrictive condition $c\,\beta^{5/7} \le {\cal O}(1)$ is enough.} Within that subspace, we were able to compute explicitly the spectrum of each of the three sectors. The results are similar to the mass spectrum of the flavor sector, found in \cite{ASW}. Most significantly, we did not find a parametrically light mode that could be identified with a dilaton.\footnote{This is in agreement with the bottom-up walking models studied in \cite{Kiritsis}.} This conclusion was also supported by the consideration of a susy probe D5 brane in the above color background. Namely, from the DBI action for that probe one can read off the couplings of the various background fluctuations with the color field-strength squared. As we saw, the composite scalar, that couples as expected for a dilaton, is not a mass eigenstate.

The restricted parameter space we considered here was very useful for building intuition and developing techniques to study this model analytically. Clearly, however, this is just the beginning regarding the exploration of the full $(c, \beta)$ parameter space of the model. It is still possible that one could find a dilaton in other corners of that parameter space, perhaps for $c \,\beta >\!\!>1 $. We will investigate this in a future publication. On the other hand, if the lack of a dilaton persists in the rest of the parameter space, that could lead to interesting conclusions based on comparison with the numerical works \cite{ENP,EP2}, which do find a light state (that is a candidate for a dilaton). Namely, the different outcomes could be partly due to a technical difference and partly to a deep conceptual one. The technical difference is that we work in the approximation $c>\!\!>1$, which is not the regime in which they find the light state, as pointed out in \cite{EP2}. The conceptual difference is that we only study the walking region, viewing it as an effective description valid only below a certain physical scale, provided by the upper end of walking.\footnote{I.e., we do not view the UV region of that background as an appropriate UV completion.} On the other hand, \cite{ENP,EP2} view their background as a UV complete description.\footnote{The cut-off introduced in parts of \cite{EP2}, in order to compare with us, is not really the same as it is not fixed by the end of walking. Also, the light state, that they find in that regime, occurs for a rather small value of the cutoff. Our considerations, on the other hand, are in a regime of large walking and thus also a large value of the cutoff. So there is no contradiction with \cite{EP2} in the present work.} Thus, it is possible that the existence of the dilaton is due to the entire background, as opposed to the walking region alone. This would suggest a rather important role for UV physics in the determination of (at least a part of) the low-energy spectrum.

We should also recall that the model, studied in \cite{LA,ASW,ASW2}, treats the flavor sector in a probe approximation. Hence, our gravity dual encodes a gauge theory, which is in a different universality class compared to the field-theoretic models of \cite{WalkT}, as already pointed out in \cite{ASW}. So this is yet another reason to be cautious as to how general our conclusion, regarding the lack of a dilaton, is. To obtain a gravitational dual of the models of \cite{WalkT}, one would like to backreact the flavor branes. This is a very difficult technical problem that, clearly, merits consideration. It should be pointed out, though, that the results of the present paper seem to indicate that taking into account the flavor backreaction would be unlikely to remedy the lack of a dilaton. Indeed, regardless of the precise mass spectrum in a backreacted model, it is natural to expect that the would be dilaton will turn out to be an even more involved combination of other fluctuations, than the one we found in Section \ref{ED} here. In other words, it is rather likely that it will not have a well-defined mass in that case as well. Leaving aside the question of flavor backreaction, one should note that our model is under better analytical control compared to the purely field-theoretic ones. So our result may have deep implications for understanding the behaviour of walking gauge theories, in particular for settling the question of whether in the walking regime the gauge theory can be well-approximated as a theory with spontaneously broken conformal symmetry or not.

Finally, it is worth noting that the mass of the lightest glueball mode in this model could be not much larger than the one for the state observed at the LHC. Additional effects, that in principle need to be taken into account, could lower our result for this mass. One such effect, according to \cite{FFS}, is the coupling to Standard Model fields. Another is the explicit consideration of a nonabelian flavor group with an appropriate number of flavors.

\section*{Acknowledgements}

We thank T. Appelquist, P. Argyres, B. Holdom, D. Hong, E. Poppitz and R. Shrock for conversations. We are also grateful to M. Piai and T. ter Veldhuis for very useful correspondence. LCRW thanks the organizers of the workshop on the origin of mass at NORDITA, P. di Vecchia, F. Sannino, K. Tuominen, C. Kouvaris and C. Pica, for their hospitality and discussions. Research at Perimeter Institute is supported by the Government of Canada through Industry Canada and by the Province of Ontario through Ministry of Research \& Innovation. L.A. is supported in part by funding from NSERC. The research of P.S. and R.W. is supported by DOE grant FG02-84-ER40153.

\appendix

\section{Derivatives of the potential} \label{A1}
\setcounter{equation}{0}

To zeroth order, i.e. evaluated on the walking background, we have:
\be
V |_w = \frac{1}{4} e^{- 4 p_w - 4 x_w} - e^{2p_w - 2 x_w} + \frac{N_c^2}{64} e^{8p_w -2x_w + \phi_w} \qquad ,
\ee
as well as:
\bea \label{Vder}
V_a &=& \frac{\pd V}{\pd a} \bigg|_w = 0 \qquad , \qquad V_b = 0 \qquad , \qquad V_g = 0 \qquad , \nn \\
V_{\phi} &=& \frac{N_c^2}{64} \,e^{8 p_w - 2x_w + \phi_w} = \frac{64 \times 3^{4/3} N_c^2}{A^5 c^6 \beta^{4/3}} \,e^{-\frac{16 \rho}{3}} \qquad , \nn \\
V_x &=& 2 e^{2p_w - 2x_w} - e^{-4p_w - 4x_w} - \frac{e^{8 p_w - 2x_w + \phi_w} N_c^2}{32} \qquad , \nn \\
V_p &=& - 2 e^{2p_w - 2x_w} - e^{-4p_w - 4x_w} + \frac{e^{8 p_w - 2x_w + \phi_w} N_c^2}{8} \qquad .
\eea
Similarly, the second derivatives of the potential, evaluated on the background, are:
\bea \label{Vderder}
V_{aa} &\equiv & \pd^2_a V |_w = - e^{2p_w - 2 x_w} + \frac{1}{2} e^{8 p_w} + \frac{e^{8 p_w - 2 x_w + \phi_w} N_c^2}{16} \quad , \nn \\
V_{ab} &\equiv & \pd_a \pd_b V |_w = -\frac{e^{8p_w - 2x_w + \phi_w} N_c^2}{16} \quad , \quad V_{bb} = \frac{e^{8 p_w -2 x_w + \phi_w} N_c^2}{32} \quad , \nn \\
V_{ag} &=& 0 \,\,\, , \,\,\, V_{ax} = 0 \,\,\, , \,\,\, V_{a\phi} = 0 \,\,\, , \,\,\, V_{ap} =0 \,\,\, , \nn \\
V_{bg} &=& 0 \,\,\, , \,\,\, V_{bx} = 0 \,\,\, , \,\,\, V_{b\phi} = 0 \,\,\, , \,\,\, V_{bp} =0 \,\,\, , \nn \\
V_{gg} &=& e^{-4p_w - 4 x_w} - e^{2 p_w - 2 x_w} + \frac{e^{8p_w - 2 x_w + \phi_w} N_c^2}{16} \,\,\, , \nn \\
V_{g\phi} &=& 0 \,\,\, , \,\,\, V_{g x} =0 \,\,\, , \,\,\, V_{g p} = 0 \,\,\, , \nn \\
V_{\phi \phi} &=& \frac{e^{8p_w - 2x_w + \phi_w} N_c^2}{64} \quad , \quad V_{\phi x} = - \frac{e^{8p_w - 2x_w + \phi_w} N_c^2}{32} \quad , \quad V_{\phi p} = \frac{e^{8p_w - 2x_w + \phi_w} N_c^2}{8} \,\,\, , \nn \\
V_{xx} &=& \frac{e^{8p_w - 2x_w + \phi_w} N_c^2}{16} - 4 e^{2 p_w - 2x_w} + 4 e^{-4p_w - 4x_w} \,\,\, , \nn \\
V_{xp} &=& - \frac{e^{8p_w - 2x_w + \phi_w} N_c^2}{4} + 4 e^{2 p_w - 2x_w} + 4 e^{-4p_w - 4x_w} \,\,\, , \nn \\
V_{pp} &=& e^{8p_w - 2x_w + \phi_w} N_c^2 - 4 e^{2 p_w - 2x_w} + 4 e^{-4p_w - 4x_w} \,\,\, .
\eea

\section{Mass Condition for $\varphi$ system} \label{MassCondPhi}
\setcounter{equation}{0}

Writing out the six equations arising from (\ref{match1}), we have:
\bea \label{6eqs}
e^{i \eta} (\alpha^* - 1 ) H_W^0 + e^{-i \eta} (\alpha - 1 ) H_W^{0*}+ d_1 H_1^0- c_1 L_{\nu}^0 &=& 0 \nn \\
e^{i \eta} (\alpha^* - 1 ) H_W^1 + e^{-i \eta} (\alpha - 1 ) H_W^{1*}+ d_1 H_1^1- c_1 L_{\nu}^1 &=& 0 \nn \\
\alpha^* e^{i \eta} H_W^0 + \alpha e^{-i \eta} H_W^{0 *}-\frac{1}{4} c_1 L_{\nu}^0 - c_2 L_0^0 &=& 0 \nn \\
\alpha^* e^{i \eta} H_W^1 + \alpha e^{-i \eta} H_W^{1 *}-\frac{1}{4} c_1 L_{\nu}^1 - c_2 L_0^1 &=& 0 \nn \\
-e^{i \eta} H_W^0 - e^{-i \eta} H_W^{0*} - \frac{1}{4} c_1 L_{\nu}^0 - c_3 L_0^0 &=& 0 \nn \\
-e^{i \eta} H_W^1 - e^{-i \eta} H_W^{1*} - \frac{1}{4} c_1 L_{\nu}^1 - c_3 L_0^1 &=& 0 \,\, .
\eea
Now, we can solve the last four equations for the four quantities $c_{1,2,3}$ and $\eta$. Writing only the results that will be needed in the following, namely the answers for $\eta$ and $c_1$, we have:
\bea
e^{2i\eta} &=& - \frac{(1+\alpha) \,(L_0^0 H_W^{1*} - L_0^1 H_W^{0*})}{(1+\alpha^*) \,(L_0^0 H_W^1 - L_0^1 H_W^0)} \label{et} \\
c_1 &=& - 4 e^{-i \eta} \,\frac{(\alpha^* - \alpha) \,(L_0^1 H_W^{0*} - L_0^0 H_W^{1*})}{(1+\alpha^*) \,(L_0^1 L_{\nu}^0 - L_0^0 L_{\nu}^1)} \, \label{c11}.
\eea
In fact, the easiest way of obtaining the above result for $c_1$ is to solve algebraically for $c_{1,2,3}$ and $e^{i \eta}$, while viewing $e^{-i \eta}$ as independent.

We can also solve algebraically the first two equations in (\ref{6eqs}) for $c_1$ and $d_1$. We find:
\be \label{c12}
c_1 = \frac{e^{i \eta} (1 - \alpha^*) (H_1^1 H_W^0 - H_1^0 H_W^1) + e^{- i \eta} (1-\alpha) (H_1^1 H_W^{0*} - H_1^0 H_W^{1*})}{H_1^0 L_{\nu}^1 - H_1^1 L_{\nu}^0} \, .
\ee
Since we will not need $d_1$, we will not write it down.

Now let us equate the right hand sides of (\ref{c11}) and (\ref{c12}):
\bea \label{MC}
&&- 4 \,\frac{(\alpha^* - \alpha) \,(L_0^1 H_W^{0*} - L_0^0 H_W^{1*})}{(1+\alpha^*) \,(L_0^1 L_{\nu}^0 - L_0^0 L_{\nu}^1)} \nn \\
&&= \frac{e^{2 i \eta} (1 - \alpha^*) (H_1^1 H_W^0 - H_1^0 H_W^1) + (1-\alpha) (H_1^1 H_W^{0*} - H_1^0 H_W^{1*})}{H_1^0 L_{\nu}^1 - H_1^1 L_{\nu}^0} \, .
\eea
Finally, substituting (\ref{et}), the condition for the mass spectrum (\ref{MC}) becomes:
\bea
&&0 = 4 ( \alpha^* - \alpha ) (L_0^1 H_W^{0*} - L_0^0 H_W^{1*}) (H_W^1 L_0^0 - H_W^0 L_0^1) (H_1^0 L_{\nu}^1 - H_1^1 L_{\nu}^0) \nn \\
&&- (1 + \alpha) (1 - \alpha^*) (L_0^0 H_W^{1*} - L_0^1 H_W^{0*}) (H_1^1 H_W^0 - H_1^0 H_W^1) (L_0^1 L_{\nu}^0 - L_0^0 L_{\nu}^1) \nn \\
&&+ (1 - \alpha) (1 + \alpha^*) (H_1^1 H_W^{0*} - H_1^0 H_W^{1*}) (H_W^1 L_0^0 - H_W^0 L_0^1) (L_0^1 L_{\nu}^0 - L_0^0 L_{\nu}^1) \, .
\eea

\end{document}